# Experimental demonstration of a scalable room-temperature quantum battery

**Authors:** Kieran Hymas[1], Jack B. Muir[1], Daniel Tibben[2], Joel van Embden[2], Tadahiko Hirai[1], Christopher J. Dunn[1], Daniel E. Gómez[2], James A. Hutchison[3], Trevor A. Smith[3], and James Q. Quach[1*]

**Affiliations:**

[1]Commonwealth Scientific and Industrial Research Organisation (CSIRO), Clayton, Victoria 3168, Australia

[2]School of Science, RMIT University, Melbourne, Victoria 3000, Australia

[3]ARC Centre of Excellence in Exciton Science, School of Chemistry, The University of Melbourne, Parkville, Victoria 3052, Australia

*Corresponding author. Email: james.quach@csiro.au

**Abstract:** Harnessing quantum phenomena in energy storage systems offers an opportunity to introduce a new generation of batteries with quantum-enhanced performance. Until now, the quantum battery has largely remained a theoretical concept, with little progress towards experimental realisation, due to the challenges in quantum coherent control. Here, we experimentally demonstrate a scalable room-temperature quantum battery with a multi-layered organic-microcavity design. We show that it exhibits superextensive charging, metastabilisation of stored energy, and generates superextensive electrical power, the latter an unpredicted phenomenon. The combination of these properties in a single device is the first demonstration of the full cycle of a quantum battery, laying the framework for future designs.

## Introduction

Driven by the transition to renewable energies and the electrification of things, it is expected that our demand for energy storage will increase by an order of magnitude over the next decade[1]. Current battery technology is unlikely to sustainably meet this demand, inviting the need for radical solutions. Quantum batteries, by operating on a different set of physical laws than conventional batteries, offers such a radical solution[2,3]. Utilising the quantum properties of entanglement, superposition, non-locality, or collective effects, they are theorised to exhibit exotic properties such as superextensive power scaling[4,5,6,7] increased bounds on energy density[8] and storage lifetime[9]. Although there is a large body of literature on the theory of quantum batteries[10-28] very few experiments exist[29,30,31,32]. These experiments investigate isolated properties, none of which demonstrate a functioning quantum battery.

The properties of physical systems can be classified as intensive, meaning they are independent of the system's size (like density), or extensive, meaning they scale with the system's size (like mass). Quantum batteries exhibit superextensive charging powers that scale faster than their size, resulting in batteries with larger capacity taking less time to charge - a property radically different from conventional electrochemical cells. This property may be driven by entanglement that minimises the number of traversed states during charging[5], or through collective effects that increase the effective coupling between the battery and its energy source[33]. For a quantum battery consisting of $N$ two-level quantum systems, this effective coupling scales as $\sqrt{N}$, so that the charging time scales inversely as $1/\sqrt{N}$. Also known as superabsorption, this property was recently demonstrated with cold-atoms[29] and organic-molecules in cavity[30]. However, these systems critically suffer from fleeting energy retention, with no mechanism for useful energy extraction.

Here, we engineer a superabsorbing quantum battery that is wirelessly charged with a laser and outputs superextensive electrical power. The quantum battery is constructed with a multi-layered microcavity design (Fig. 1a). The microcavity is tuned to the resonant frequency of the ground to first-excited singlet transition of the absorber molecule, copper phthalocyanine (CuPc), to induce a strong light-matter coupling. We utilise this strongly coupled singlet excitation to rapidly charge the battery, and a weakly coupled triplet state to store the energy. The charging laser is tuned to the lower polaritonic state, that arises from the entanglement of the absorber singlet states with the cavity mode. Using ultrafast spectroscopy, we show that the charging rate superextensively scales with the number of absorber molecules. Upon charging, the energy is quickly transferred to the metastable triplet state, increasing the storage time to six orders of magnitude longer than the charging time. This allows a sufficient period for the stored energy to be extracted as electrical work. Electrical extraction is facilitated by charge transport layers to affect an energy gradient for charge separation and transport, with additional blocking layers to prevent recombination. The result is an electrical power output that scales superextensively with the capacity of the battery. By incorporating superextensive charging, metastable energy storage, and superextensive electrical power output in a single device, our work lays out a framework for practical quantum batteries.

## Results

**Device engineering**

We fabricated eight quantum battery devices (labelled D1 to D8) based on Fabry-Perot microcavities[34] whose optical length was tuned to the absorption frequency of an ensemble of CuPc molecules. The number of CuPc absorbers $N$ was adjusted in each device to span $N \approx 2.8 \times 10^{14}$ for the smallest battery to $N \approx 7.9 \times 10^{14}$ for the largest battery. We will show that our device is scalable up to $6 \times 10^{10}$ superabsorbing molecules, limited only by the laser spot size. To facilitate energy extraction from each device via a charge current we incorporated a charge donor-acceptor combination of CuPc and fullerene ($C_{60}$)[35,36,37] as well as a suitable choice of optically inert electron-hole blocking and transport materials that were integrated into the cavity[38]. A schematic of the device is shown in Fig. 1A where we also indicate the direction of laser irradiation. For each device, we fabricated two identical structures without the top and bottom silver mirrors, to act as optical and electrical controls, respectively (see Fig. S2). To minimise parameter variability the controls were fabricated on the same substrate and fabrication run (see Materials and Methods). In Fig. 1B we schematically depict the work functions and HOMO/LUMO energy levels for each layer of the cavity device. The layer composition induces an energy gradient within the cavity that facilitates charge separation of excitons created by superabsorption in the CuPc layer. The layer composition was chosen to maximise electrical performance of the quantum battery without drastically detuning the fundamental frequency of the cavity from the CuPc Q-band.

Each CuPc molecule can effectively be described as a non-interacting four level system (Fig. 3B). The first excited singlet $S_1$ is separated from the ground state $S_0$ by ~ 2 eV with a lower-lying triplet state $T_1$ at approximately 1.2 eV[39]. The excited singlet features result from Davydov splitting, common to most phthalocyanine absorbers[40]. This feature is prominent in the reflection measurements of the optical no-cavity devices (black curve in Fig. 1C) and manifests as two weakly split signals with approximately equal oscillator strengths. In the presence of a cavity, these absorption bands are strongly hybridised with the fundamental mode of the confined photon field leading to the formation of upper (UP), middle (MP), and lower (LP) polariton states (red curve in Fig. 1C). By systematically varying the concentration of CuPc molecules in the mixed CuPc:$C_{60}$ layer while keeping the cavity length fixed, we were able to tune the Rabi splitting between the polariton branches as a function of $N$. As a consequence of adjusting the CuPc concentration and layer thicknesses, the fundamental cavity frequency did not remain constant across the devices, leading to shifting asymmetries in the UP and LP reflectance bands (Fig. S3).

The reflectance spectra as a function of incidence angle for the quantum battery is shown in Fig. 1D. Overlaid are the theoretically predicted triple polaritonic states. In comparison, Fig. 1E shows the reflectance spectra of the no-cavity control, which is overlaid with two Davydov split states.

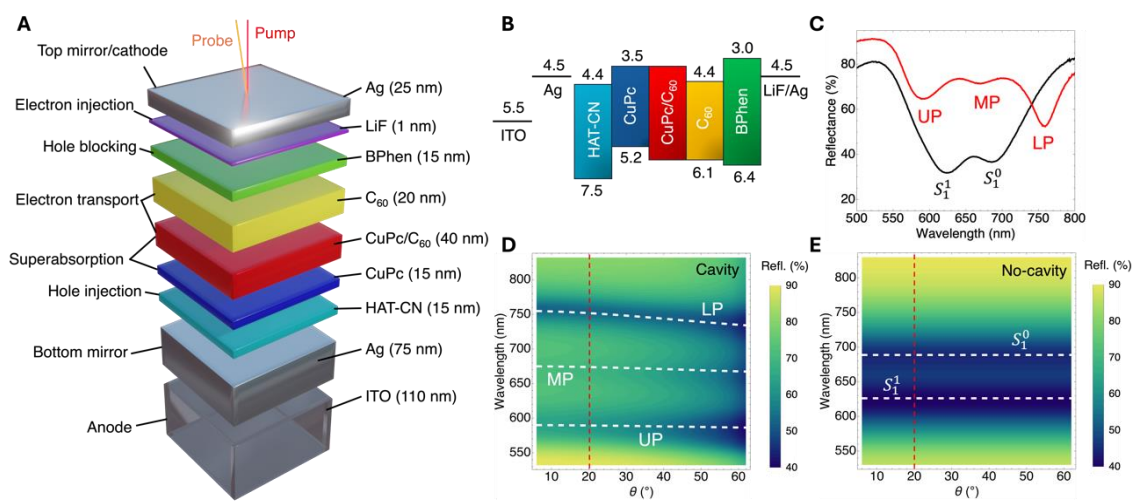

**Fig. 1. Composition of the quantum battery tuned for strong light-matter coupling.** (**A**) Schematic of the layered structure of the quantum battery, describing the function and composition of each component. Ultrafast pump and probe laser pulses are used to charge and measure the device. (**B**) Work functions and HOMO/LUMO energy levels (in units of eV) of each layer of the quantum battery, defining an energy gradient for charge separation. (**C**) Steady-state reflectance spectrum at 22° incidence angle for the D5 quantum battery (red line) and the optical control device (black line), which has no top mirror. The control spectrum is characterised by singlet states $S_1^0$ and $S_1^1$. In the quantum battery, these singlet states are entangled with the photonic cavity mode to give rise to polaritons, as characterised by the upper (UP), middle (MP), and lower (LP) polariton states. (**D**) and (**E**) show the reflection spectra for the quantum battery and control as a function of incidence angle; the dashed red line corresponds to the reflectance spectra shown in (C).

**Superextensive charging with photons**

In this section we demonstrate the superextensive charging properties of our quantum battery devices. To reveal their dynamics, we utilise ultrafast transient spectroscopy; a technique that probes electronic and vibrational evolution in molecules and semiconductors on femtosecond time scales[41]. In our experiments, a laser pump pulse resonant with the LP branch induces excitations in the quantum battery, whose time evolution is monitored by a delayed probe pulse resonant with the UP branch. Using this scheme, we are able to monitor the temporal dynamics of the excited state populations (see Materials and Methods).

The observed excited state population of a representative quantum battery device is shown in Fig. 2A (orange squares), where each data point is an average of 25,000 measurements. Superimposed on these experimental data points is the theoretical total excited population of CuPc, which is calculated by summing the population of the first excited singlet states ($S_1^0, S_1^1$) and the triplet state ($T_1$) (see Materials and Methods). There is good agreement between the experimental data and the theoretically predicted populations. Using the known energy values of the singlet and triplet states[36], we calculate the total energy of the system from their excited populations (green line in Fig. 2A).

For each device, Fig. 2B summarises the estimated number of CuPc absorbers $N$ in the pump laser cross section, the maximum stored energy per absorber ($E_{max}$), the time to reach half maximum energy beginning from $1/e^2$ of the Gaussian laser pulse ($\tau$), and the peak charging power density ($P_{max} = E_{max}/\tau$). These values are plotted in Figs. 2C to 2E. Changes in the composition of the devices as $N$ was varied, led to slight differences in the cavity resonances and Q factors (supplementary Text). In Figs. 2C to 2E, we also plot the theoretical predictions for a device with cavity resonance ($\Delta_c = 1.87$ eV) and loss ($\kappa = 33$ ps$^{-1}$) averaged over the eight devices. Due to these variances, the experimental data points slightly deviate from the plot of a theoretical device with averaged cavity properties. In particular, D8 deviates the most from these average values, since this device exhibits the largest detuning from the Davydov resonances, explaining its relatively low $E_{max}$ and $P_{max}$ (which fall out of the range of the Fig. 2C and 2E plots). Nevertheless, we clearly see a superextensive increase in the charging power and energy density, accompanied with a subextensive charging time.

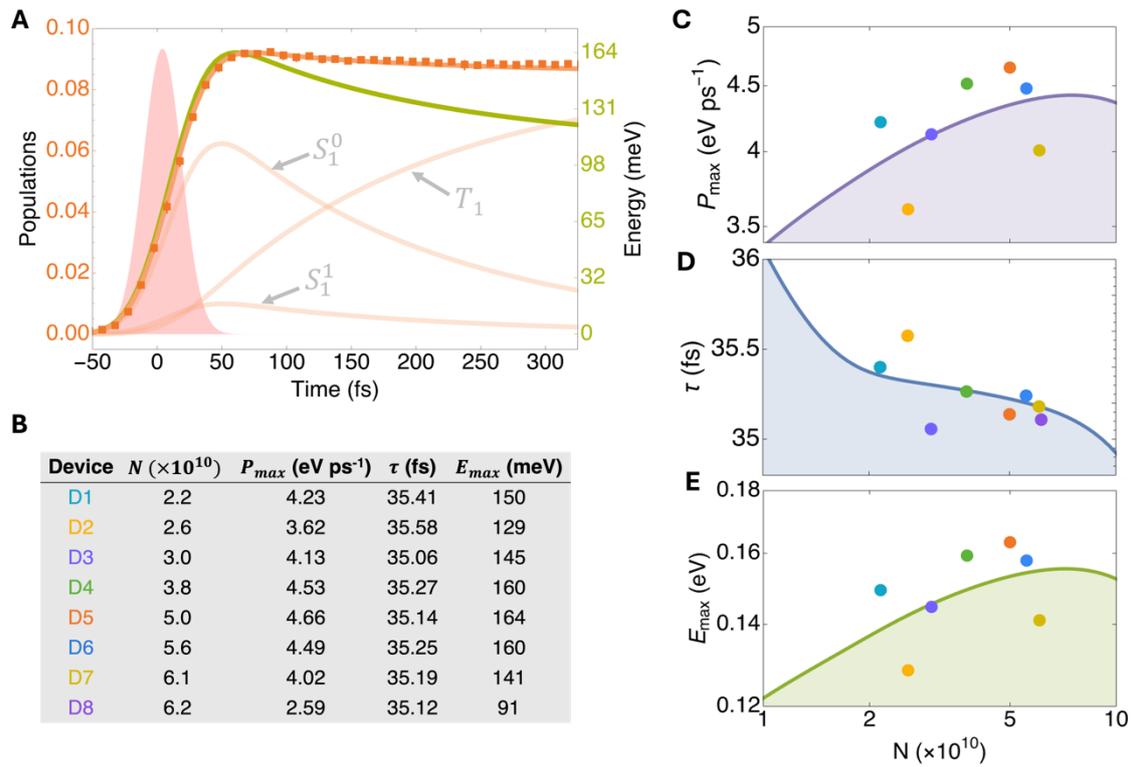

**Fig. 2. Superextensive charging dynamics.** (**A**) A laser pulse excites the CuPc molecules. The orange squares are the measured excited molecule populations for a representative device. (1$\sigma$ standard deviation error bars are less than the dot size). Overlaid (orange line) is the theoretical prediction, calculated by summing the first excited singlet states ($S_1^0, S_1^1$) and the triplet state ($T_1$) populations, convoluted with the instrument response function. The green line gives the average total energy per molecule. (**B**) Table of device name, number of CuPc molecules $N$, peak power density $P_{max}$, charging time $\tau$, and maximum energy per molecule $E_{max}$. These experimental data points are plotted in (**C**, **D**, **E**) following the colour scheme in (B). The uncertainty in $N$ is less than 8%, which is smaller than the dot size. The solid lines in (C, D, E) plot $P_{max}$, $\tau$ and $E_{max}$ as a function of $N$ for an idealised device with cavity frequency $\Delta_c = 1.87$ eV and cavity loss $\kappa = 33$ ps$^{-1}$, which are the average values of the eight experimental devices. The devices show a superextensive charging power and energy density, since these properties increase with $N$. The charging time is subextensive, as it decreases with $N$.

**Stabilisation of stored energy**

In this section we demonstrate the metastability of the stored energy in our quantum battery devices. We measure their long-time decay dynamics using a spectrally broad supercontinuum probe with spectral width spanning 500-1000 nm. The transient reflectance spectra for all devices are reported in Fig. S12, with a representative spectrum of D5 given in Fig. 3A. The positive (red) signals in the transient reflectance spectra result from ground state bleaching; and the negative (blue) differential reflectance signal, results from excited state absorption. Each cavity device exhibits differential reflectance signals that persist for tens of nanoseconds, showing that the excited state remains populated for six orders of magnitude longer than the device charging time. These long-excited state lifetimes originate from intramolecular relaxation of the singlet excitations to metastable triplet excitations as shown in Fig. 3B. At approximately 200 fs, intersystem crossing (ISC) becomes the dominant singlet relaxation mechanism forming metastable triplet excitations (purple line in Fig. 3C) with typical lifetimes on the order of 10-50 ns. This is three orders of magnitude longer than the state-of-the-art in quantum batteries[30]. The efficiency of this conversion mechanism is enhanced by spin-orbit coupling of the $Cu^{2+}$ central metal ion that works to invert the spin of the excited phthalocyanine electron. Thus, the system is blocked from relaxing from the triplet $T_1$ to the singlet ground state $S_0$ by Pauli exclusion, resulting in energy metastabilisation in the quantum battery.

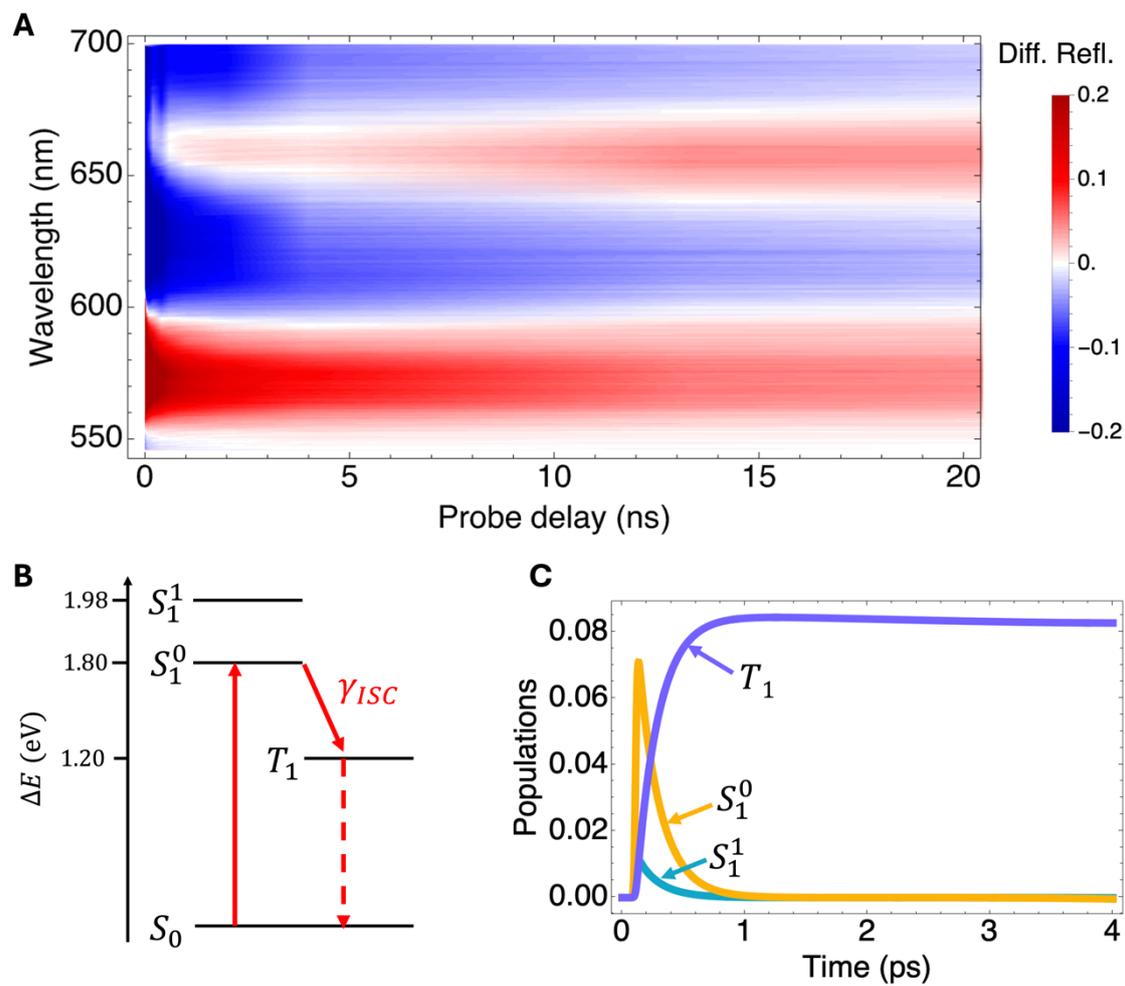

**Fig. 3. Metastabilisation of stored energy via intersystem crossing.** After the collective excitation of the singlet states during the charging process, the energy is stored in the metastable triplet state via fast intersystem crossing. (**A**) Transient differential reflectivity of device D5. The persistence of the signals indicates that excited states remain populated for tens of nanoseconds, which is six orders of magnitude longer than the device charging time. (**B**) Energy splitting of relevant states of the CuPc absorber with dominate absorption and relaxation pathways highlighted as red arrows. The fluorescence quantum yield of CuPc is approximately zero due to rapid intersystem crossing $\gamma_{ISC}^{-1} = 200$ fs. (**C**) Time-evolution of the mean excited singlet ($S_1^0$ yellow and $S_1^1$ blue) and triplet ($T_1$ purple) populations of the CuPc: within approximately 200 fs, rapid intersystem crossing converts singlet excitons into metastable triplet excitons.

**Superextensive discharging of electrical power**

Controlled extraction of stored energy is a necessary feature of batteries. To power a small electronic device, for example, it is desirable for the extracted energy to be available as electrical work. Here we demonstrate the enhanced electrical energy output of our quantum batteries. In Figs. 4A and 4B we plot the ratio of ejected electrons to incident photons, or external quantum efficiency (EQE), for the cavity and no-cavity controls. A comparison of the two figures shows that the cavity devices have a three-fold enhancement over the no-cavity controls. The polariton branches are also visible in the EQE measurements of the cavity.

In addition to the EQE, we investigated the discharging power of the devices. Fig. 4C plots current-voltage (I-V) curves for the cavity and no-cavity control of a representative device D5 (red lines), shown with their steady-state discharging power (blue lines). The optimal operating point for energy extraction is at the peak of the discharge power curve. We investigate the ratio of cavity to no-cavity peak discharging power, $P_c^{max}/P_{nc}^{max}$; as the cavity and no-cavity controls were fabricated on the same substrate, this ratio accounts for variations between fabrication runs. In Fig. 4D we observe an increase in $P_c^{max}/P_{nc}^{max}$ with $N$, indicating superextensive discharging power in the devices. (Note that we consider device D3 to be an overperforming outlier.) In comparison, we point out that a constant discharging power ratio greater than one and independent of $N$, would indicate a non-superextensive enhancement. This superextensive discharging power has never been observed, and suggests a collective enhancement stemming from the strong light-matter coupling in the quantum batteries. Multidimensional coherent spectroscopy[42] and ultrafast photocurrent measurements[43] offer potential avenues to further interrogate the microscopic origin of this enhancement.

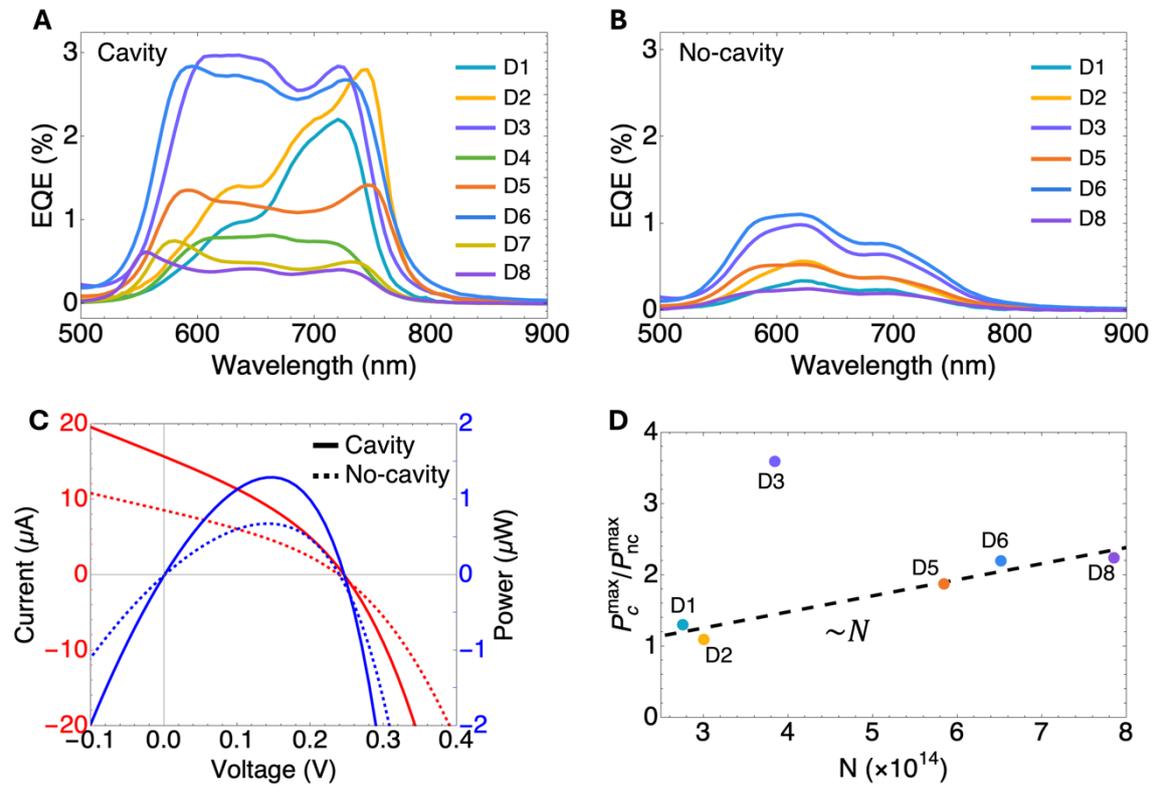

**Fig. 4. Energy extraction from the quantum battery as electrical work.** A comparison of the external quantum efficiency of the (**A**) quantum batteries and their (**B**) control devices, show a threefold photon to charge conversion enhancement in the former. (**C**) Current (red lines) and discharging power (blue lines) shown as a function of voltage for a representative device D5 (solid lines) and its control (dotted lines). (**D**) plots the ratio of the peak discharging power of the quantum batteries and their control devices. This ratio increases with *N*, indicating that the discharge power of the quantum battery scales superextensively. Note: the control devices for D4 and D7 were not fabricated with electrical controls and thus are not reported here.

## Discussion

Until now, the study of quantum batteries has been predominately a theoretical endeavour with scarce experimental verification. However, through finely tuned fabrication and systematic spectroscopic analysis we have experimentally demonstrated a full operational cycle of a quantum battery, from superextensive charging to enhanced electrical energy extraction. Our device operates at room temperature, and we have demonstrated scalability of up to $6 \times 10^{10}$ superabsorbing molecules.

By fine-tuning the device we were able to demonstrate superextensive charging in the strong light-matter coupling regime: the device benefitted from strong dephasing mechanisms which operated on a timescale only slightly longer than the superabsorption, to quench superradiant self-discharge. At longer timescales, intersystem crossing efficiently shuttles singlet excitations into metastable triplet states with lifetimes as long as tens of nanoseconds, before dissociating into free carriers and ultimately contributing to a photocurrent. We have demonstrated with quantum efficiency measurements that our quantum batteries capture photons and convert their energy into useful electrical work more efficiently than no-cavity controls. Future studies could employ time-resolved XPS measurements as a direct probe of the charge separation in CuPc:$C_{60}$ complexes[44]. Using such a technique may provide a pathway to explore superextensive photocurrents that leverage polaron- polaritons[45].

As the first demonstration of a full quantum battery operational cycle, our devices lay the foundation for future energy storage technologies that utilise quantum effects for improved performance. How to scale the energy density of these technologies remains an open question that must be resolved to compete with conventional electrochemical cells.

Arranging microcavity quantum batteries in planar arrays that share a common electrode affords a promising direction to achieve this. Although our quantum battery was charged by a coherent laser source, this work also opens a pathway for charging with incoherent sunlight, offering an exciting new approach to the design of solar-cell technology. Besides the development of practical technologies, by coupling collective excitations to the generation of charge currents, our device also acts as a platform for exploring foundational photophysics at the frontier of ultrafast light-to-charge conversion.

## Methods

### Device fabrication

ITO-coated glass substrates (Asahi Glass Co. Ltd. Japan) were cleaned with isopropyl alcohol and deionized water in an ultrasonic bath. The substrates were then subjected to a UV−ozone treatment in a UV−ozone cleaner (Novascan PSD Pro) to obtain an oxygen-rich ITO surface, increasing the work function of ITO. A 75 nm thick Ag (Sigma-Aldrich Solutions) bottom mirror layer was then deposited on the ITO layer, followed by a 15 nm hole-injection HAT-CN (Luminescence Technology Co.) layer under high vacuum. From then on, all layers up to the Ag top mirror layer were deposited using a thermal evaporation method under high vacuum ($<1.0\times10^{-4}$ Pa). Next, a 15 nm CuPc layer, a mixed layer of CuPc/$C_{60}$ (Luminescence Technology Co.), and a $C_{60}$ layer were sequentially deposited (the thicknesses of these layers varied for each device). A 15 nm BPhen (Luminescence Technology Co.) and a 1 nm LiF (Sigma-Aldrich Solutions) layer were then deposited as electron transport and injection layers. Finally, a 25 nm Ag (Sigma-Aldrich Solutions) top mirror layer was deposited on the LiF layer. The devices were encapsulated with cover glass and a desiccant, and then sealed with a UV-curable epoxy resin. The final device is ~$2 \times 2$ cm$^2$ and has four separated $2 \times 5$ mm$^2$ regions of cavity, and two $2 \times 5$ mm² regions of no cavity. Further fabrication details can be found in the Supplementary Information.

**Steady-state spectroscopy**

Angle–resolved reflectometry measurements were performed with an Agilent Cary 7000 UV-Visible-NIR spectrophotometer with Universal Measurement Accessory (UMA) and xenon lamp source. Measurements across the devices were baseline-corrected and performed with standardised spot size and lamp intensity. Fixed-angle reflectometry and transmission measurements were performed with an Agilent Cary 5000 UV-Visible-NIR spectrophotometer with Diffuse Reflectance Accessory (DRA) and xenon lamp source. Measurements across the devices were baseline-corrected and performed with standardised spot size and lamp intensity. Reflectance measurements were collected at 8° angle of incidence to enable collection of specular and diffuse reflectance signal.

**Ultrafast spectroscopy**

We use femtosecond transient reflection spectroscopy to probe the depletion of a common ground state between the LP and UP and circumvent pump scatter which typically obscures pump-probe measurements. The pump-probe measurement produces a differential reflectivity

$$\frac{\Delta R}{R} = \frac{R_{ON} - R_{OFF}}{R_{OFF}}$$

where $R_{ON}$ ($R_{OFF}$) is the probe reflectivity with (without) a pump pulse. Since an increase in probe reflectivity in response to the pump is associated with the quantum system absorbing energy (i.e., by inducing a resonant transition $i \to f$ in the quantum battery), $\Delta R/R$ is proportional to the population $P_i$ of the initial state $i$[30]. We choose a probe wavelength resonant with the ground to upper polariton transition such that an increase in differential reflectivity provides a time-resolved measure of the ground state population or, conversely, of the sum of all excited state populations.

Both two-colour pump-probe and pump-supercontinuum-probe measurements were performed on the quantum battery devices in reflection geometry, at room temperature, in air. The two-colour femtosecond laser pulses were generated by two NOPAs (Light Conversion, Orpheus-N-2H and Orpheus-N-3H). The NOPAs were pumped by a Yb:KGW laser amplifier (Light Conversion Pharos, 1030 nm, 180 fs, 100 µJ pulses). A secondary identical amplifier is seeded by the oscillator of the first amplifier (Pharos Duo, Light Conversion). The output of the secondary amplifier was focused onto a 2 mm sapphire crystal to generate a broadband supercontinuum which spanned 500-1000 nm. Amplifier pulses were selected at a repetition rate of 33.3 kHz for all measurements. The two-colour pump-probe was used to measure the sub-100 fs dynamics, while the pump-supercontinuum-probe was used to measure spectrally broad long-time dynamics.

We determined and optimised the pulse durations using auto- and cross-correlation measurements (Figs. S6 and S7). The pump pulses from the 2H-NOPA had an average pulse width (Gaussian FWHM) of 34±1 fs. The probe pulses from the 3H-NOPA had an average pulse width of 32±3 fs across all measurements. We used the measured cross-correlation widths (~50 fs) for the instrument response function in the theoretical model. The pump and probe beams were focused onto the QB devices with an almost collinear geometry (~4° between beam paths) with a FWHM spot diameter of 32 µm for the pump, and 42 µm for the probe. We have checked that the relative sizing between the pump and probe does not change the dynamics of the $\Delta R/R$ signal. The QB devices were rotated 25° to the incoming pump and probe pulses so the reflected probe beam could be spatially isolated and re-collimated. The probe was then directed into a spectrometer with a high-speed CCD (Entwicklungsbuero Stresing, FLC3030) triggered by the laser amplifier. To measure the $\Delta R/R$ signal, the pump was modulated by a mechanical chopper operating at 16.67 kHz (SciTec Instruments, 310CD. To scan the $\Delta R/R$ signal as a function of time, the probe beam was sent through optical retroreflector delay lines, which extended to 8 ns for the supercontinuum probe pulses (Newport, DL325), or to 0.8 ns for the NOPA probe pulses (Newport, ESP300). Longer supercontinuum probe delay times out to ~16 µs were achieved by electronically triggering the secondary amplifier from later oscillator pulses (period of 13.3 ns) relative to the first amplifier.

Since the chopper spins at half the repetition rate of the laser, sequential probe-only and pump-probe pulses are incident on the devices, thus enabling the acquisition of shot-to-shot $\Delta R/R$. The two-colour pump probe measurements were scanned over the probe delay of -5 to 10 ps, with -200 to 400 fs in steps of 10 fs. At each probe delay time, 5000 shot-to-shot $\Delta R/R$ spectra are averaged. Measurements across the entire probe delay were repeated 5 times under identical experimental conditions, and the average of these measurements were taken, with the 1σ standard deviation as the error bars.

To compare the ultrafast dynamics across different devices it was important to keep a constant ratio of incident photons to absorber molecules in the laser volume. To achieve this, the fluence and wavelength of the pump-probe pulses were varied to account for the steady-state reflection spectra of each device (see Supplementary Information). The fluence of the pump as a function of $\Delta R/R$ intensity followed a linear dependence for all fluences used in the experiments (see Supplementary Information), ensuring higher order effects induced by the pump are not present in the measurements.

**External quantum efficiency measurements**

External quantum efficiency of the devices was calculated as the ratio of the number of photons absorbed by the device to the number of photons that participate in generating a photocurrent. This is given by $\hbar\omega I/eP$, where $I$ is the photocurrent, $\hbar\omega$ is photon energy, $P$ is the power of incident photons, and $e$ is electron charge.

External quantum efficiency spectroscopy was performed at the *RMIT Nanostructures Laboratory* using a PV Measurements QEXL quantum efficiency instrument with Xenon arc lamp and monochromator, calibrated with a Si photodiode at 300-1100 nm. Internal quantum efficiency was calculated by dividing the experimentally measured external quantum efficiency by the absorption at each wavelength, obtained by fixed–angle UV-Visible-NIR reflectance and transmission spectroscopy on an Agilent Cary 5000 UV-Vis-NIR spectrophotometer with DRA and xenon arc lamp.

**Discharging power measurements**

I-V measurements were collected at the *RMIT Nanostructures Laboratory* using a Keithley 2636B SYSTEM SourceMeter® and collimated 625 nm LED source (ThorLabs M625L4) at 10 mW/cm², with a resolution of 0.01 V. Measurements proceeded by first illuminating the device with the LED source, then varying the bias voltage from 1.5 V to

-1.5 V. The power of each device was calculated at the point where the product of current and voltage is at a maximum, i.e., the maximum power point, according to P = IV.

**Theoretical model**

The reflectance spectra were modelled using a coupled oscillator Hamiltonian

$$H_{CO} = \Delta_c a^\dagger a + \Delta_1 X_{11} + \Delta_2 X_{22} + g_{co}(a^\dagger X_{01} + X_{10} a + a^\dagger X_{02} + X_{20} a) \quad (1)$$

where $a^{(\dagger)}$ annihilates (creates) excitations of the confined photon field with frequency $\Delta_c$ and $X_{\alpha\beta} = |\alpha\rangle\langle\beta|$ are Hilbert operators that work on the low-lying electronic levels of the CuPc absorber. The indices $\alpha, \beta \in \{0,1,2\}$ enumerate the ground $S_0$, first-excited $S_1^0$ and second-excited $S_1^1$ levels, respectively, with energies 0, $\Delta_1$ and $\Delta_2$ as in Fig. 3a. Since we do not probe the steady-state reflectance of our devices in the near infra-red, the triplet state dynamics do not contribute to the simulated reflectance, thus we do not consider their population in our coupled oscillator model. The collective light-matter coupling constant $g_{co}$ quantifies the interaction of the CuPc ensemble electric dipole with the quantised cavity radiation field and is related to the bare coupling $g$ via $g_{co} = \sqrt{N} g$. The steady-state reflectance of the cavity was simulated within a Fermi golden rule approach

$$R \approx 1 - A = 1 - I_0 \sum_\mu |\langle \phi_\mu | a^\dagger | \phi_0 \rangle|^2 e^{-(\epsilon_\mu - \epsilon_0 - \nu)^2 / 2\sigma} \quad (2)$$

involving the coupled oscillator eigenstates $|\phi_\mu\rangle$ with corresponding energies $\epsilon_\mu$ that depend upon the microscopic parameters of the Hamiltonian (1). The intensity $I_0$ and combined homogeneous and inhomogeneous broadening $\sigma$ are fitting parameters and $\nu$ is the frequency of incident photons external to the quantum battery.

The pertinent dynamics of the ultrafast charging experiments are captured by the Tavis-Cummings Hamiltonian

$$H(t) = (\Delta_c - \nu)a^\dagger a + i\eta(t)(a^\dagger - a) \tag{3}$$

$$+ \sum_{j=1}^{N} \left((\Delta_1 - \nu)X_{11}^{(j)} + (\Delta_2 - \nu)X_{22}^{(j)} + \Delta_T X_{TT}^{(j)}\right)$$

$$+ g \sum_{j=1}^{N} \left(a^\dagger X_{01}^{(j)} + X_{10}^{(j)} a + a^\dagger X_{02}^{(j)} + X_{20}^{(j)} a\right)$$

written in the rotating frame of the laser. We assume a Gaussian temporal profile $\eta(t) = \exp\left[-((t - t_0)/2\sigma)^2\right]/\sigma\sqrt{2\pi}$ for the pulse envelope and FWHM of 34 fs. The dissipative interactions between the cavity quantum battery and its environment are accounted for with the Lindblad master equation

$$\dot{\rho} = -i[H(t), \rho] + \kappa \mathcal{L}[a] + \sum_{j=1}^{N} \left(\gamma^- \left(\mathcal{L}\left[X_{01}^{(j)}\right] + \mathcal{L}\left[X_{02}^{(j)}\right]\right) + \gamma_T^- \mathcal{L}\left[X_{0T}^{(j)}\right]\right. \tag{4}$$

$$\left. + \gamma^z \left(\mathcal{L}\left[X_{11}^{(j)}\right] + \mathcal{L}\left[X_{22}^{(j)}\right]\right) + \gamma_{\text{ISC}} \left(\mathcal{L}\left[X_{T1}^{(j)}\right] + \mathcal{L}\left[X_{T2}^{(j)}\right]\right)\right)$$

where $\mathcal{L}[O] = O\rho O^\dagger - \frac{1}{2}\{O^\dagger O, \rho\}$ are Lindblad jump operators. In Eq. (4) we account for the rate of photon loss from the cavity $\kappa$, the rate of non-radiative relaxation from the singlet $\gamma^-$ and triplet $\gamma_T^-$ states, intersystem crossing $\gamma_{\text{ISC}}$ and singlet dephasing $\gamma^z$.

The ultrafast transient reflectance experiments provide a measure of the summed excited state CuPc populations $\Delta R/R \propto \langle X_{TT} \rangle + \langle X_{11} \rangle + \langle X_{22} \rangle$. The populations $\langle X_{\alpha\alpha} \rangle$ for a molecule in the ensemble are obtained by numerically solving a hierarchical set of coupled equations of motion for the molecular and cavity degrees of freedom which, in general, do not close. We enforce the closure of these equations by approximating the time-evolution of three body correlators as $\langle ABC \rangle = \langle AB \rangle \langle C \rangle + \langle AC \rangle \langle B \rangle + \langle A \rangle \langle BC \rangle - 2\langle A \rangle \langle B \rangle \langle C \rangle$, thus capturing the pertinent dynamics with corrections scaling as $O(1/N)$.

By solving for the populations $\langle X_{\alpha\alpha}\rangle$, we obtain the time-evolution of the quantum battery energy density $E(t) = \hbar(\Delta_1\langle X_{11}\rangle + \Delta_2\langle X_{22}\rangle + \Delta_T\langle X_{TT}\rangle)$, the charging time $\tau$ (defined by $E(\tau) = E_{\max}/2$) and the maximum charging power $P_{max} = E_{\max}/\tau$.

## Acknowledgements

J.A.H. and T.A.S. acknowledge financial support through various Australian Research Council (ARC) schemes; LE200100051, CE170100026, FT180100295 and infrastructure support through the Australian Centre for Advanced Photovoltaics (ACAP) Infrastructure Scheme. K.H. and J.Q.Q. acknowledge funding from the Revolutionary Energy Storage Systems Future Science Platform.

## Author Contributions

J.Q.Q. conceived and managed the project. K.H. and J.Q.Q. developed the theoretical model and analysis. D.T., J.vE., and D.G. performed and interpreted steady-state reflectance and electrical measurements on the devices. J.B.M. and T.A.S. contributed to the ultrafast measurements. T.H. and C.J.D. fabricated the devices. All authors contributed to discussion of the results and the writing of the manuscript.

## Competing interests

The authors declare no competing interests.

*Supplementary Information:* Experimental demonstration of a scalable room-temperature quantum battery

Kieran Hymas,[1] Jack B. Muir,[1] Daniel Tibben,[2] Joel van Embden,[2] Tadahiko Hirai,[1] Christopher J. Dunn,[1] Daniel E. Gómez,[2] James A. Hutchison,[3] Trevor A. Smith,[3] and James Q. Quach[1, *]

[1]*Commonwealth Scientific and Industrial Research Organisation (CSIRO), Clayton, Victoria 3168, Australia*
[2]*School of Science, RMIT, Melbourne, Victoria 3000, Australia*
[3]*ARC Centre of Excellence in Exciton Science, School of Chemistry, University of Melbourne, Parkville, Victoria 3052, Australia*

**CONTENTS**

* [james.quach@csiro.au](mailto:james.quach@csiro.au)

# I. DEVICE COMPOSITION AND GEOMETRY

## A. Device schematics

| D1 | D2 | D3 | D4 | D5 | D6 | D7 | D8 |
|---|---|---|---|---|---|---|---|
| Glass top | Glass top | Glass top | Glass top | Glass top | Glass top | Glass top | Glass top |
| Ag 25nm | Ag 25nm | Ag 25nm | Ag 25nm | Ag 25nm | Ag 25nm | Ag 25nm | Ag 25nm |
| LiF 1nm | LiF 1nm | LiF 1nm | LiF 1nm | LiF 1nm | LiF 1nm | LiF 1nm | LiF 1nm |
| BPhen 15nm | BPhen 15nm | BPhen 15nm | BPhen 15nm | BPhen 15nm | BPhen 15nm | BPhen 15nm | BPhen 15nm |
| $C_{60}$ 75nm | $C_{60}$ 70nm | $C_{60}$ 40nm | $C_{60}$ 60nm | $C_{60}$ 40nm | $C_{60}$ 40nm | $C_{60}$ 40nm | $C_{60}$ 40nm |
| CuPc:$C_{60}$ 30%:70% 5nm | CuPc:$C_{60}$ 30%:70% 10nm | CuPc:$C_{60}$ 20%:80% 40nm | CuPc:$C_{60}$ 70%:30% 20nm | CuPc:$C_{60}$ 50%:50% 40nm | CuPc:$C_{60}$ 60%:40% 40nm | CuPc:$C_{60}$ 70%:30% 40nm | CuPc:$C_{60}$ 80%:20% 40nm |
| CuPc 15nm | CuPc 15nm | CuPc 15nm | CuPc 15nm | CuPc 15nm | CuPc 15nm | CuPc 15nm | CuPc 15nm |
| HAT-CN 15nm | HAT-CN 15nm | HAT-CN 15nm | HAT-CN 15nm | HAT-CN 15nm | HAT-CN 15nm | HAT-CN 15nm | HAT-CN 15nm |
| Ag 75nm | Ag 75nm | Ag 75nm | Ag 75nm | Ag 75nm | Ag 75nm | Ag 75nm | Ag 75nm |
| ITO 110nm | ITO 110nm | ITO 110nm | ITO 110nm | ITO 110nm | ITO 110nm | ITO 110nm | ITO 110nm |
| Glass Substrate | Glass Substrate | Glass Substrate | Glass Substrate | Glass Substrate | Glass Substrate | Glass Substrate | Glass Substrate |

**Fig. S1**. **Composition and geometry of each device.** Devices are shown with increasing number of absorbers $N$ from left to right. While the physical length between the mirrors is fixed at 126 nm, the optical length varies slightly due to the different refractive index profiles shifting the cavity resonance between devices.

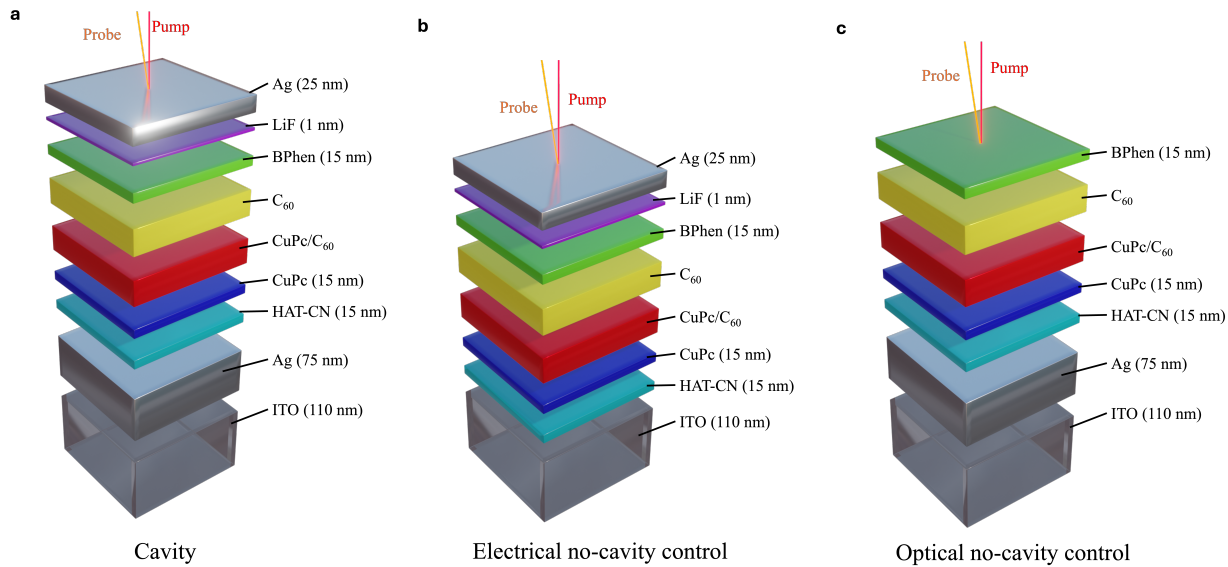

**Fig. S2**. **Exploded view of cavity and no-cavity controls.** Each device is fabricated with four cavities and two no-cavity controls. Of the two no-cavity controls, we fabricate an electrical no-cavity that is missing the 75 nm Ag layer (i.e. the bottom mirror) and an optical no-cavity that is missing the 25 nm Ag layer (i.e. the top mirror).

## B. Further fabrication details

The devices were fabricated on patterned ITO-glass (8 $\Omega$ sq$^{-1}$, Asahi Glass Co. Ltd. Japan) by vacuum evaporation of layer materials through a set of shadow masks in a thermal evaporator (Kitano-Seiki Co. Ltd. Japan) in high vacuum ($< 1.0 \times 10^{-4}$ Pa) with mask changes occurring without fully breaking vacuum. The layer materials Dipyrazino[2,3-f:2',3'-h]quinoxaline-2,3,6,7,10,11-hexacarbonitrile (HAT-CN), Copper (II) Phthalocyanine (CuPc), Fullerene-C$_{60}$ (C$_{60}$) and 4,7-Diphenyl-1,10-phenanthroline (BPhen) were purchased from Luminescence Technology Co., Silver (Ag) and Lithium Fluoride (LiF) were purchased from Sigma-Aldrich Solutions. All materials were used as received after degassing. Deposition tooling factors were determined prior to device fabrication for all materials by profilometer measurements (Bruker Dektak 150) on a 50 nm thick film of each layer material. Each layer was deposited at a deposition rate of 1 Å s$^{-1}$ during device fabrication, except for LiF which was deposited at 0.1 Å s$^{-1}$.

The number of absorbers $N$ was controlled by a combination of layer composition and thickness for a given fabrication cycle. The device design, that is: with the absorbers interposed between top and bottom silver layers comprising a cavity, or two types of no-cavity controls in which either one of the two silver layers forming the cavity was absent (see fig. S2), was determined by the choice of shadow mask used during silver and LiF depositions during a given fabrication cycle. This resulted in a total of six separate patterned areas on a single substrate, each of dimensions $2 \times 5$ mm$^2$; four test devices with a cavity, and one each of two types of no-cavity controls, all with identical $N$, being produced during a particular fabrication cycle. The first of these controls, lacking the thick silver bottom mirror layer, was used as a no-cavity electrical control device to which electrical contacts were made through the ITO as the bottom electrode. The second no-cavity control lacks the top electrode and was used for optical studies.

The fabrication details of device D5, with the structure depicted in fig. 1A of the main text and the corresponding no-cavity controls (as shown in fig. S2), follow as a representative description of the fabrication cycle.

Immediately prior to use, the patterned ITO-coated glass substrate (Asahi Glass Co. Ltd. Japan) was cleaned with isopropyl alcohol and deionized water in an ultrasonic bath. It was then subjected to a UV-ozone treatment in a UV-ozone cleaner (Novascan PSD Pro).

to obtain an oxygen-rich ITO surface, increasing the work function of ITO, and then rapidly transferred into the vacuum chamber operating under high vacuum ($< 1.0 \times 10^{-4}$ Pa) so that the first layer deposition commenced within $\sim$ 10 minutes of completion of the UV-ozone treatment.

First, a 75 nm thick Ag bottom mirror layer was deposited through a shadow mask onto the ITO layer in five of the six device positions on the substrate. This was followed by a change to a mask with a single large area aperture used to deposit the non-electrode materials over a single large area square across all six device positions. There followed sequential depositions under high vacuum of a 15 nm hole-injection layer of HAT-CN, a 15 nm layer of CuPc, co-deposition of 40 nm of a mixed layer of CuPc:$C_{60}$ in the volumetric ratio 50%:50%, a 40 nm $C_{60}$ layer and then 15 nm of BPhen as electron transport layer. There then followed a change to a mask for the deposition of the top electrodes comprising a 1 nm LiF electron injection layer and then a 25 nm Ag top mirror layer. This latter mask was patterned such that only four of the five identical devices with the bottom silver mirror were also capped with the top silver mirror. Substrates were then transferred to a nitrogenfilled glovebox (water and oxygen 0.3 ppm) and the devices encapsulated under a nitrogen atmosphere in the recess formed under a hollowed cover glass and containing a desiccant pad, sealed in place with a UV-curable epoxy resin applied to the border outside the device area. After curing, an encapsulated substrate could be removed from the glovebox and was handled in air during further study.

The final device was $2 \times 2$ cm$^2$ and had four separated $2 \times 5$ mm$^2$ cavity regions for testing as quantum batteries, and two different no-cavity control regions, each also of $2 \times 5$ mm$^2$. During testing, electrical contacts are made to extensions of the top and bottom electrodes that fall outside the encapsulation glass along orthogonal directions from the cavity or no-cavity electrical devices. This device fabrication cycle was repeated with the different CuPc:$C_{60}$ mixed layer compositions and thicknesses and the different $C_{60}$ layer thicknesses to give the range of $N$ values corresponding to the complete device set D1 to D8 (as shown in fig. S1) and their corresponding no-cavity controls.

## II. DEMONSTRATION OF STRONG LIGHT-MATTER INTERACTIONS

### A. Steady-state reflectance measurements

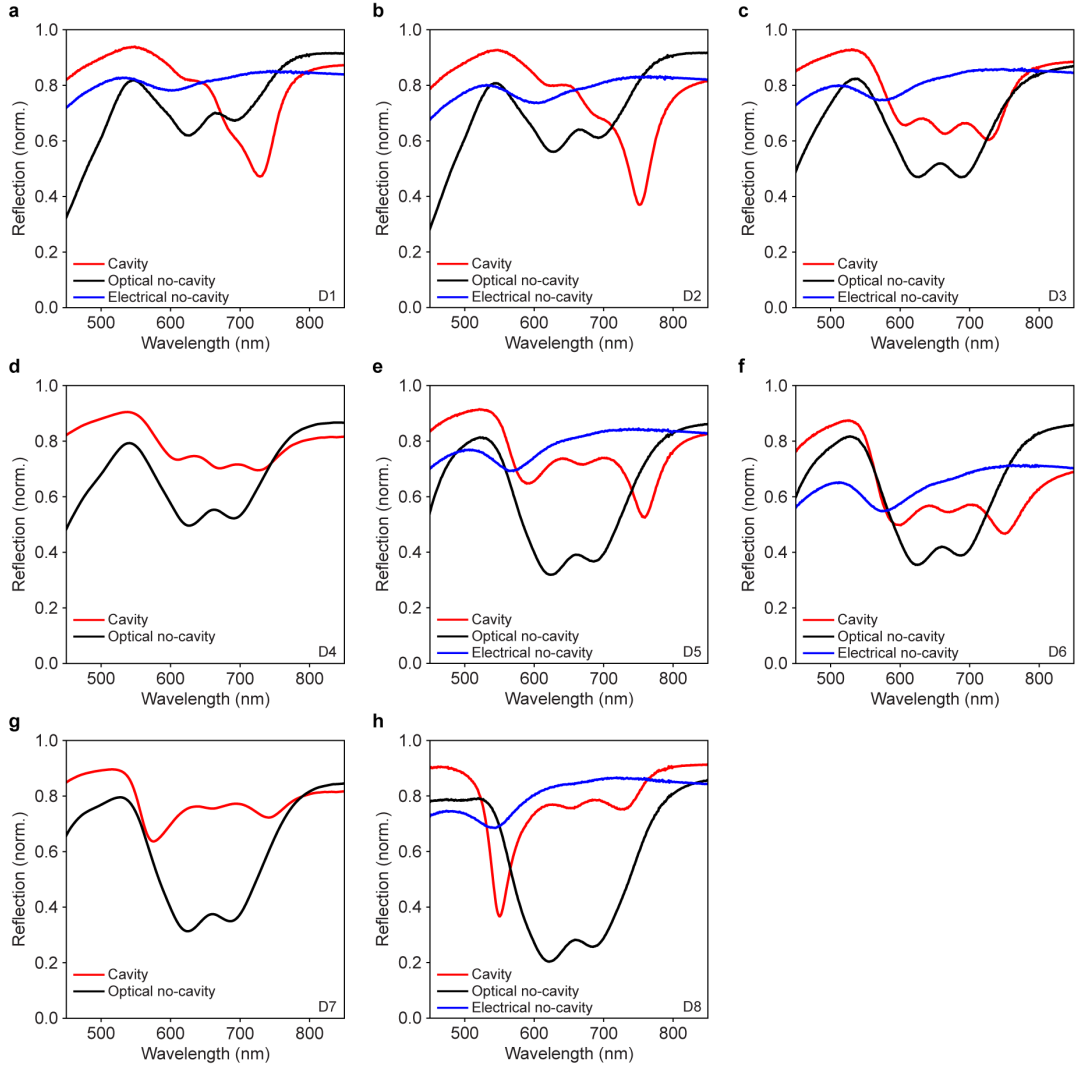

**Fig. S3**. **Steady-state reflection spectra for each device.** Devices are shown with increasing number of absorbers $N$ from **a** to **h**.

### B. Coupled oscillator model

To model the steady-state reflectance spectrum of each device, we employed a coupled oscillator model for the CuPc absorbers coupled to a single cavity mode. We diagonalised the coupled oscillator Hamiltonian

$$H_{\text{CO}} = \Delta_c a^\dagger a + \Delta_1 X_{11} + \Delta_2 X_{22} + g_{\text{co}} \left( a^\dagger (X_{01} + X_{02}) + (X_{10} + X_{20}) a \right) \quad (1)$$

where $a^{(\dagger)}$ destroys (creates) a cavity photon with energy $\Delta_c$, $X_{\alpha\beta} = |\alpha\rangle\langle\beta|$ (for $\alpha, \beta \in \{0, 1, 2\}$) is the Hilbert operator for the molecular ensemble (modelled as a single, three-level system) with Davydov split excited levels $\Delta_1$ and $\Delta_2$ and $g_{\text{co}}$ is the collective light-matter interaction. The energies $\epsilon_\mu$ and eigenstates $|\phi_\mu\rangle$ of the coupled oscillator single-excitation manifold are ready obtained by diagonalisation of Eq. (1). The simulated reflectance spectra were obtained via Fermi's golden rule

$$R \approx 1 - A = 1 - I_0 \sum_\mu \left| \langle \phi_\mu | a^\dagger | \phi_0 \rangle \right|^2 e^{-(\epsilon_\mu - \nu)^2 / 2\sigma} \quad (2)$$

where $R$ is light reflected from the cavity, $A$ is light absorbed by the cavity, $\nu$ is the frequency of the applied radiation, $I_0$ is the radiation intensity, $\sigma$ is a broadening factor and $|\phi_0\rangle = |0\rangle_{\text{ph}} \otimes |0\rangle_{\text{mol}}$ is the non-degenerate ground state of the device. Comparisons between the experimental spectra and theoretical spectra are shown in fig. S4 where the key features (lower, middle and upper polariton branches) of the reflectance spectra are captured by the theoretical curves.

In Table S1 we compile the parameters that yielded the best fits to the experimental steady-state reflectance data. $\Delta_1$ and $\Delta_2$ remained fixed, as indicated by the optical no-cavity reflectance spectra (fig. S3), and $\Delta_c$ and $g_{\text{co}}$ were allowed to vary freely. We found the collective light-matter coupling constant increased consistently across devices D1 to D8 as the number of CuPc absorbers increased in the cavity. Notably, the cavity frequency $\Delta_c$ changed between devices somewhat erratically, as a consequence of changes in the device composition and geometry. For devices D1-6, the cavity frequency is approximately tuned to the lower Davydov resonance $\Delta_1$ leading to a lower polariton (LP) branch with strong photonic character (see the deeper reflectance troughs at the LP for D1-6 in fig. S4). For D7 and D8, the cavity frequency is more closely tuned to the second Davydov resonance $\Delta_2$ leading to upper polariton (UP) branches with strong photonic character. The coupled oscillator wavefunctions for devices D1 and D8 at each polariton branch (as a function of $\Delta_c$) are shown in fig. S5.

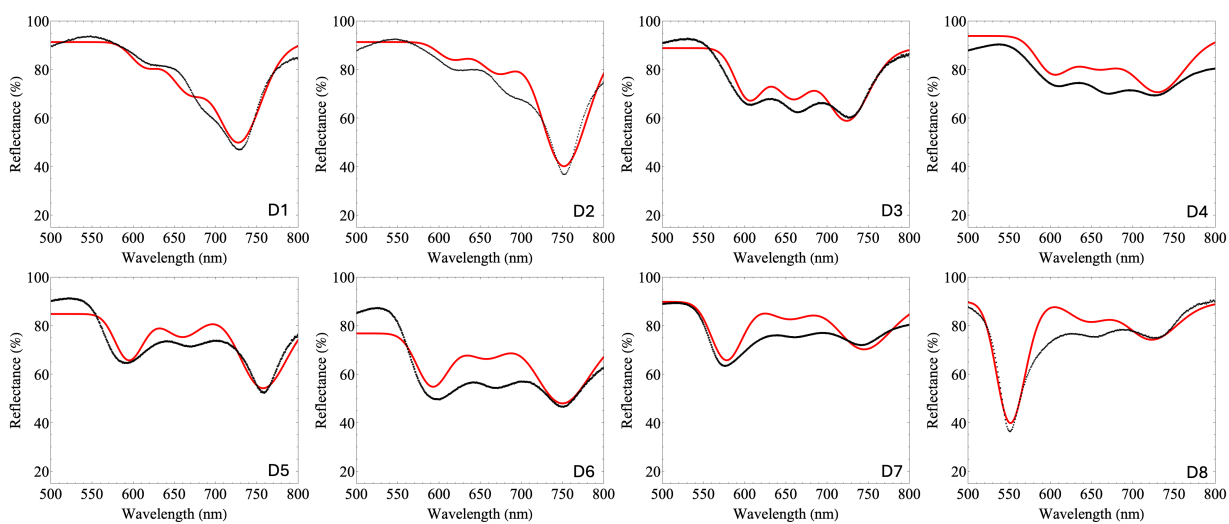

**Fig. S4**. **Comparison of experimental and simulated steady-state reflectance spectra.** Normal incidence reflection spectra of all devices D1-8 studied in this manuscript. Experimental data are shown as black dots and theoretical spectra from the coupled oscillator model are shown as solid red lines.

In a prior work, the weak signature of a middle polariton (MP) was observed in absorption and EQE measurements performed on a similar ultrastrong coupled CuPc microcavity device[1]. In that work, however, the signal was attributed to weakly allowed absorption from dark states uncoupled to the cavity mode. Notably, in our coupled oscillator model it was crucial to include coupling to the cavity mode from both Davydov resonances to reproduce the complex behaviour of the central resonance (at $\sim 650$ nm) across each device. Analysis of the coupled oscillator eigenstates for each device at $\sim 650$ nm suggested MP wavefunctions in which both Davydov resonances are hybridised with the photon field. As the cavity mode becomes strongly detuned from either Davydov resonance, however, the photonic character of the wavefunction is greatly diminished and the resonance at $\sim 650$ nm becomes significantly weaker (compare the central panels of fig. S5).

The bare light-matter coupling for a single molecule to the cavity mode can be approximated by $g = g_{\text{co}}/\sqrt{N}$ where $N$ is the number of molecules coupled to the cavity mode. We

| Device | $\Delta_1$ (eV) | $\Delta_2$ (eV) | $\Delta_c$ (eV) | $g_{co}$ (meV) | $I_0 \times 10^2$ | $\sigma \times 10^2$ (eV$^{-1}$) |
|---|---|---|---|---|---|---|
| D1 | 1.80 | 1.98 | 1.79 | 80 | 1.2 | 6.0 |
| D2 | 1.80 | 1.98 | 1.72 | 85 | 1.2 | 6.0 |
| D3 | 1.80 | 1.98 | 1.86 | 100 | 1.2 | 6.0 |
| D4 | 1.80 | 1.98 | 1.85 | 109 | 1.0 | 7.0 |
| D5 | 1.80 | 1.98 | 1.82 | 142 | 1.0 | 6.0 |
| D6 | 1.80 | 1.98 | 1.85 | 141 | 1.2 | 7.0 |
| D7 | 1.80 | 1.98 | 1.94 | 157 | 1.0 | 7.0 |
| D8 | 1.80 | 1.98 | 2.10 | 158 | 1.4 | 7.0 |

**Table. S1**. **Coupled oscillator simulation parameters.** The free parameters of the coupled oscillator model are reported above however the energy levels $\Delta_1$ and $\Delta_2$ were fixed in our simulations as evidenced by the no-cavity steady-state spectra shown in fig. S3.

approximate the number of CuPc molecules within a given volume using

$$N = \frac{N_A \rho V}{M} \tag{3}$$

where $N_A = 6.02 \times 10^{23}$ mol$^{-1}$ is Avogadro's constant, $\rho \approx 1.6$ g cm$^{-3}$ and $M = 576.07$ g mol$^{-1}$. The spot size of the impinging radiation on our device was roughly 4 mm$^2$. In Table S2 we report the calculated bare light-matter interaction and obtain an average $g = 8.37$ neV with a standard deviation of 0.73 neV.

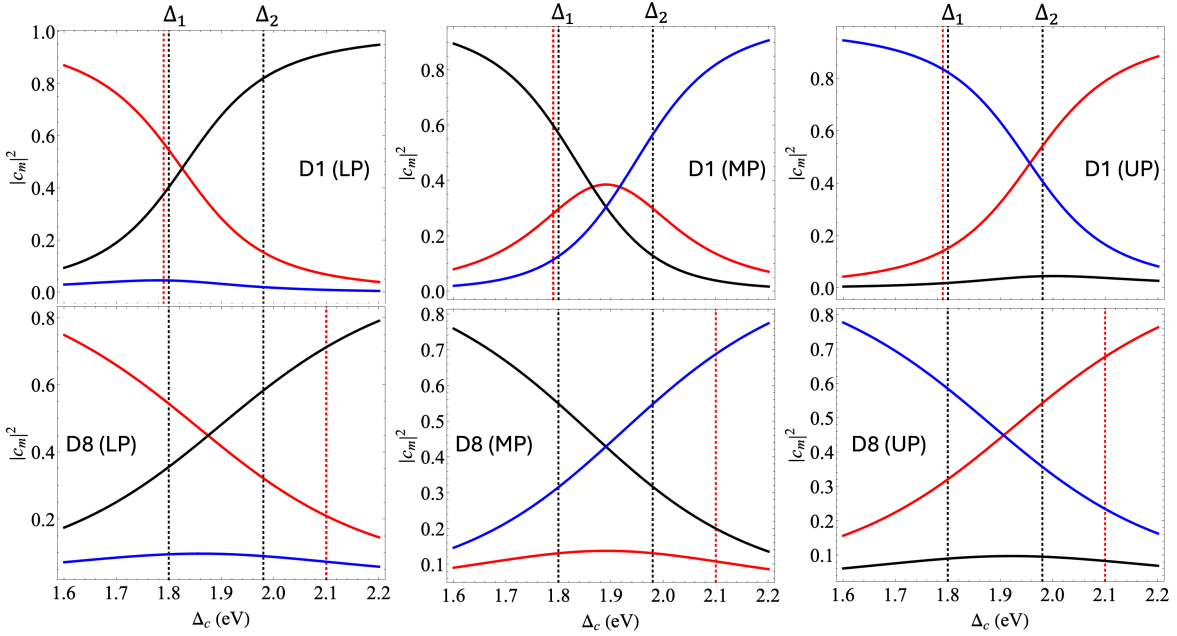

**Fig. S5**. **The effect of cavity detuning on CuPc polariton wavefunction.** The squared amplitudes of the lower, middle and upper polariton wavefunction $\left|\phi^{\mathrm{LP/MP/UP}}\right\rangle = c_1^{\mathrm{LP/MP/UP}}\left|1\right\rangle_{\mathrm{ph}} \otimes \left|0\right\rangle + c_2^{\mathrm{LP/MP/UP}}\left|0\right\rangle_{\mathrm{ph}} \otimes \left|1\right\rangle + c_3^{\mathrm{LP/MP/UP}}\left|0\right\rangle_{\mathrm{ph}} \otimes \left|2\right\rangle$ shown as a function of detuning $\Delta_c$ for D1 (top row) and D8 (bottom row). The squared amplitudes $\left|c_1^{\mathrm{LP/MP/UP}}\right|^2$, $\left|c_2^{\mathrm{LP/MP/UP}}\right|^2$ and $\left|c_3^{\mathrm{LP/MP/UP}}\right|^2$ are shown as red, black and blue solid lines, respectively. The molecular resonances $\Delta_1$ and $\Delta_2$ are shown as solid black dashed lines and the fitted cavity frequency for each device is shown as a red dashed line.

| Device | $N(\times 10^{14})$ | $g_{co}/\sqrt{N}$ (neV) |
|--------|---------------------|--------------------------|
| D1 | 1.10 | 7.63 |
| D2 | 1.20 | 7.30 |
| D3 | 1.54 | 8.03 |
| D4 | 1.94 | 7.81 |
| D5 | 2.34 | 9.30 |
| D6 | 2.61 | 8.73 |
| D7 | 2.88 | 9.25 |
| D8 | 3.14 | 8.94 |

**Table. S2**. **Approximation of the light-matter interaction strength.** We report the approximate number of CuPc molecules in the cross sectional volume of the light beam for each device using Eq. (3). The light-matter interaction strength for a single CuPc molecule coupled to the cavity mode is approximated via $g_{co}/\sqrt{N} = 8.37$ neV with a standard deviation 0.73 neV.

## III. ULTRAFAST SPECTROSCOPY

### A. Ultrafast laser pulse characterisation

We characterised the pulse duration at the sample position using (auto-) cross-correlation techniques, and subsequently compressed the pulses to account for temporal dispersion of the optics (including the device encapsulation glass). We used a 0.02 mm thick BBO crystal (United Crystals) to generate the sum frequency of the NOPA pulses, which was detected using a photodiode and lock-in synced to the laser amplifier (50 kHz). A dichroic filter and apertures were used to block the fundamental NOPA pulses while letting the sum frequency pass. An electronically controlled delay stage was used to scan one of the NOPA's pulses in time with respect to the other.

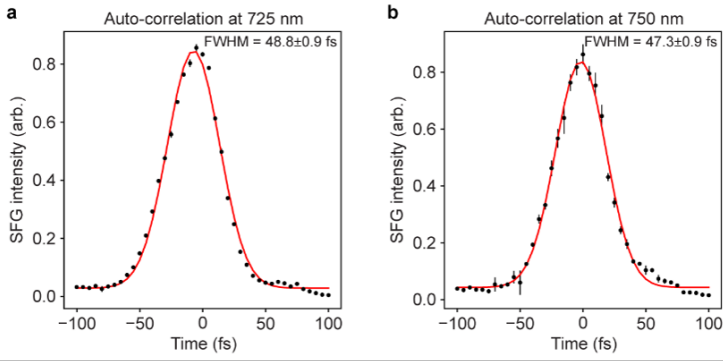

**Fig. S6**. **Auto-correlation of the pump pulses. a** sum frequency generation (SFG) intensity of two 725 nm pulses as a function of inter-pulse delay, fit with a Gaussian. The error on the FWHM fit is derived from the first standard deviation of the fit. **b** auto-correlation at 750 nm. From these measurements, we ensure the pump pulse duration is $34 \pm 1$ fs across all measurements.

The auto-correlation of the 2H NOPA ("pump") pulses was optimised to a FWHM of $48.8 \pm 0.9$ fs at 725 nm, and $47.3 \pm 0.9$ fs at 750 nm, as shown in fig. S6. Assuming a Gaussian pulse, the convoluted pulse duration can be determined by dividing by a factor of $\sqrt{2}$, which gives a FWHM of $34 \pm 1$ fs for the pump pulse duration across all experiments.

Since the devices have spectrally varying polariton absorption, measuring on resonance required different probe wavelengths between 545 and 625 nm, with the pump at 725 or 750 nm. Cross-correlations for every combination of pump and probe are shown in fig. S7. The device that uses this configuration is also shown. From the known pump pulse duration and cross correlation, we can determine the 3H NOPA ("probe") pulse duration by convolving two Gaussians.

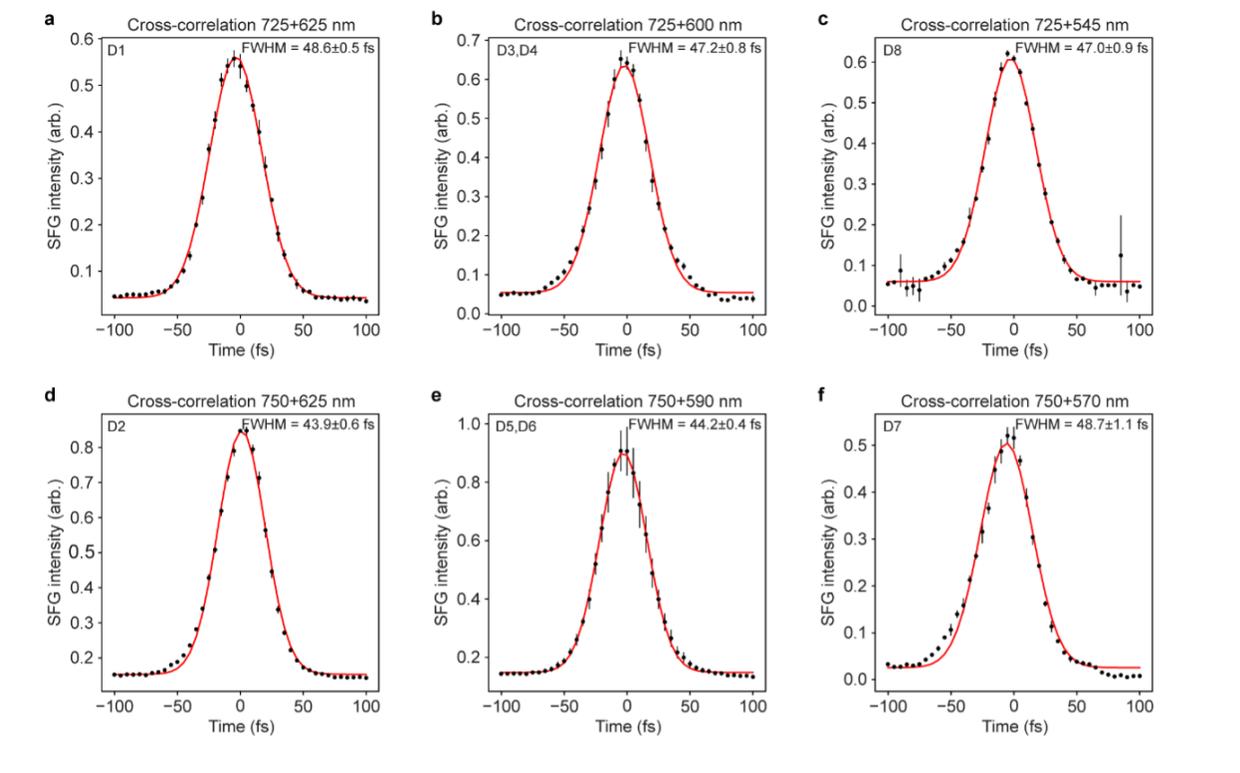

**Fig. S7**. **Cross-correlation of the pump-probe pulses. a-f**, sum frequency generation (SFG) intensity of pump and probe pulses as a function of inter-pulse delay, fit with a Gaussian. The probe pulse duration for **a**, **b**, **c**, **f** is $\sim 34$ fs, while for **d**, **e** it is $\sim 28$ fs.

### B. Power dependence of pump probe measurements

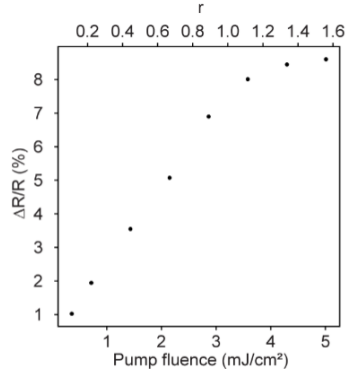

**Fig. S8**. **Power dependence of pump probe measurements.** The $\Delta R/R$ intensity of D6 as a function of 750 nm pump fluence that hits the device, with the probe fluence kept constant at $\sim 3~\mu\text{J/cm}^2$. The $r$ value ($r = \kappa N_\gamma/N$) for D6 is also displayed for reference. The $\Delta R/R$ is linear with pump fluence until $\sim 3.53$ mJ/cm$^2$, at which point it begins to saturate. All experiments are performed with $r = 0.5$ and $< 3.3$ mJ/cm$^2$ which falls in the linear regime, ensuring contributions from higher order effects are minimised.

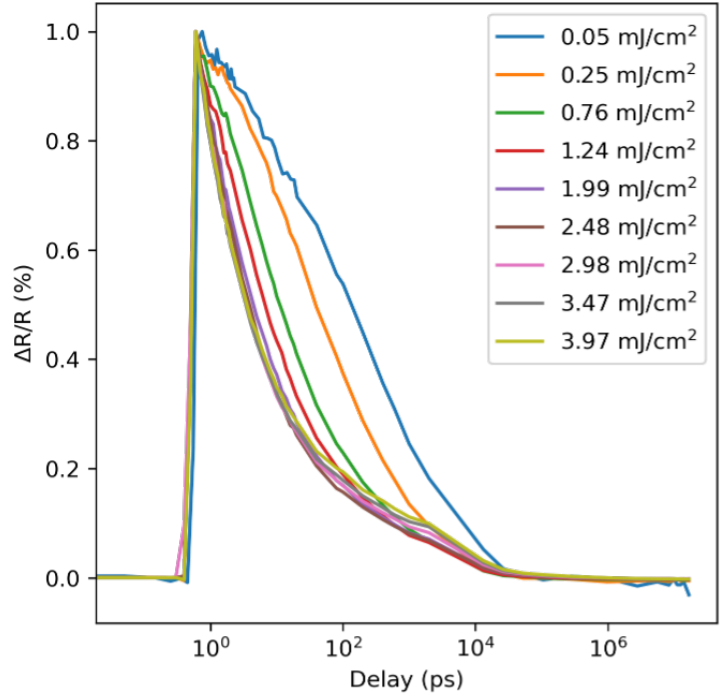

**Fig. S9**. **Long delay time power dependence.** The normalised $\Delta R/R$ intensity sliced at $\sim 625$ nm for D8, as a function of probe delay time and 725 nm pump fluence, with the probe fluence kept constant at $\sim 20$ $\mu J/cm^2$. Exciton-exciton annihilation significantly speeds up the decay as the fluence (and hence exciton density) increases. At $\sim 2$ $mJ/cm^2$ this effect saturates, and the lifetimes do not decrease any further.

## C. Ultrafast charging of the quantum battery

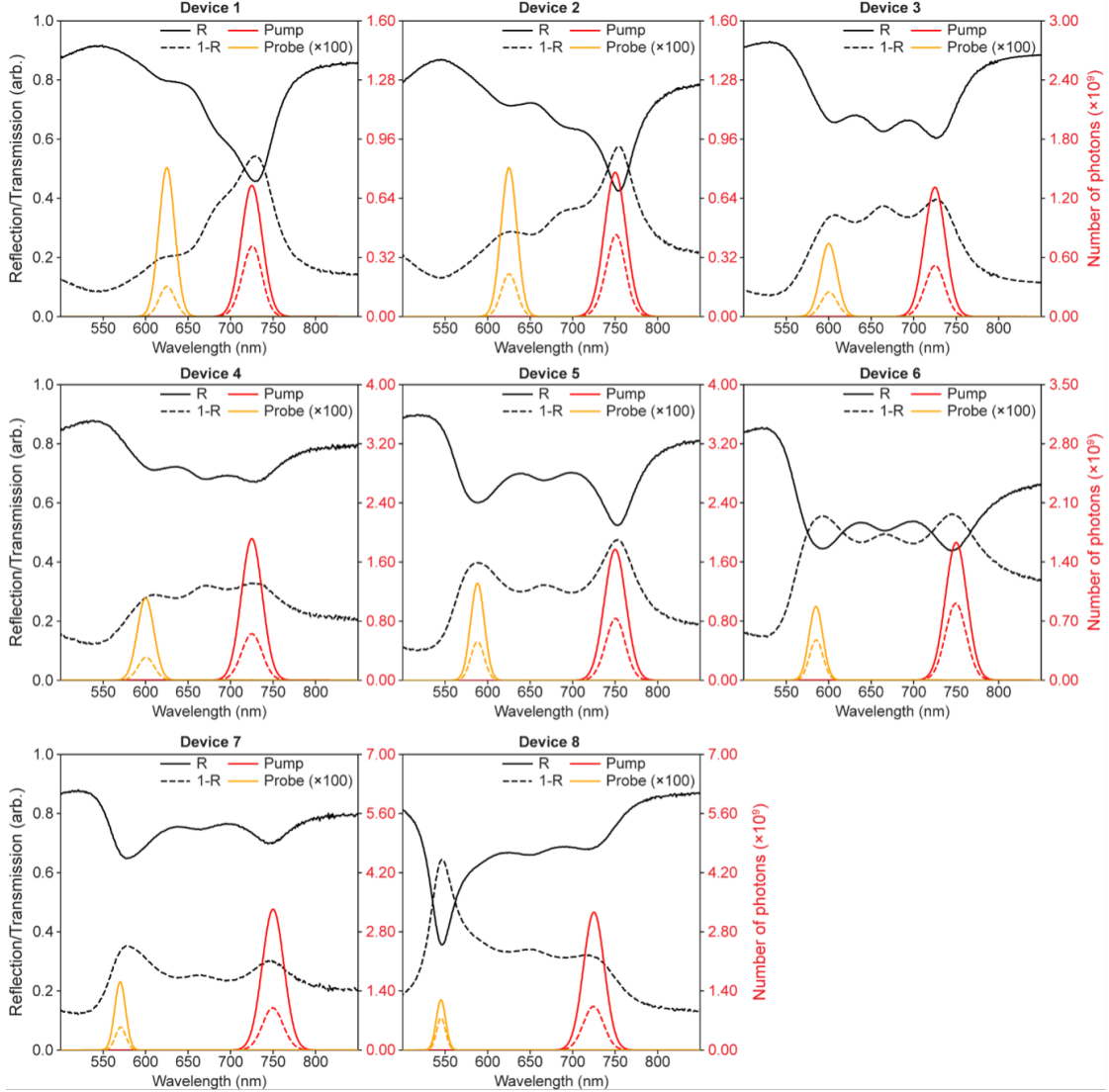

**Fig. S10**. **Pulse configuration for each quantum battery device.** Steady-state reflection of D1-D8 cavities at 25° (black lines) with pump (red Gaussian) and probe (yellow Gaussian). The pulse spectra are multiplied by 1-R (dashed line) to determine the number of pulse photons which reach the inside of the cavity (dashed Gaussians). The spectral width of the pulses is obtained from the internal NOPA spectrometers, and the number of photons is calculated using the average power, repetition rate, and spot sizes. Note, the probe pulses always have $\sim 100$ times less photons.

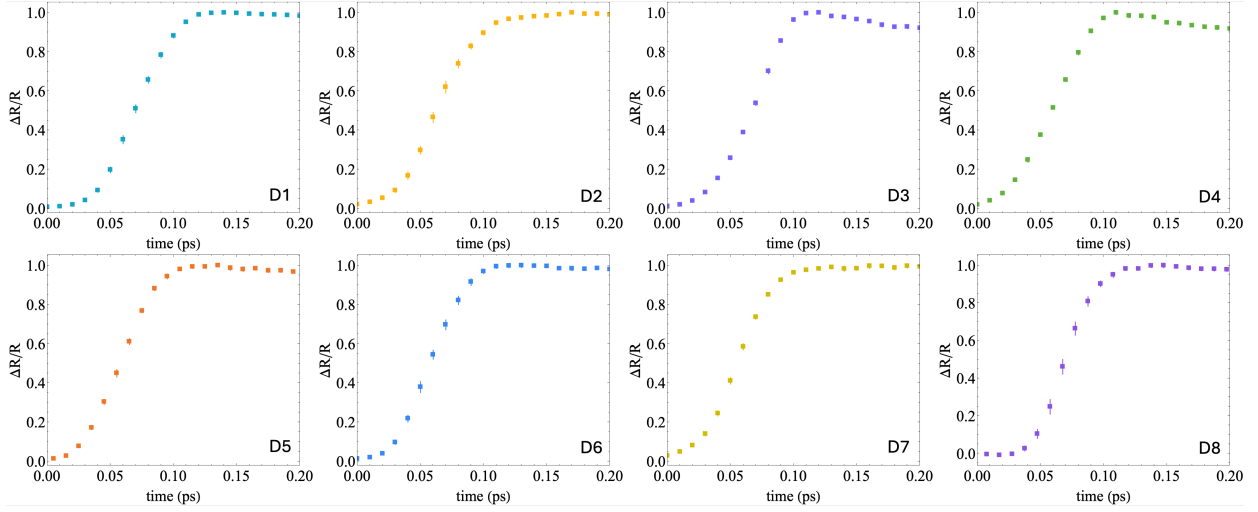

**Fig. S11**. **Ultrafast charging of each quantum battery device.** Experimental pump probe differential reflectivity of each quantum battery device as a function of time using the pump configuration specified in fig. S10. The associated error is shown as a bar superimposed with each data point.

## D. Long-time delay measurements

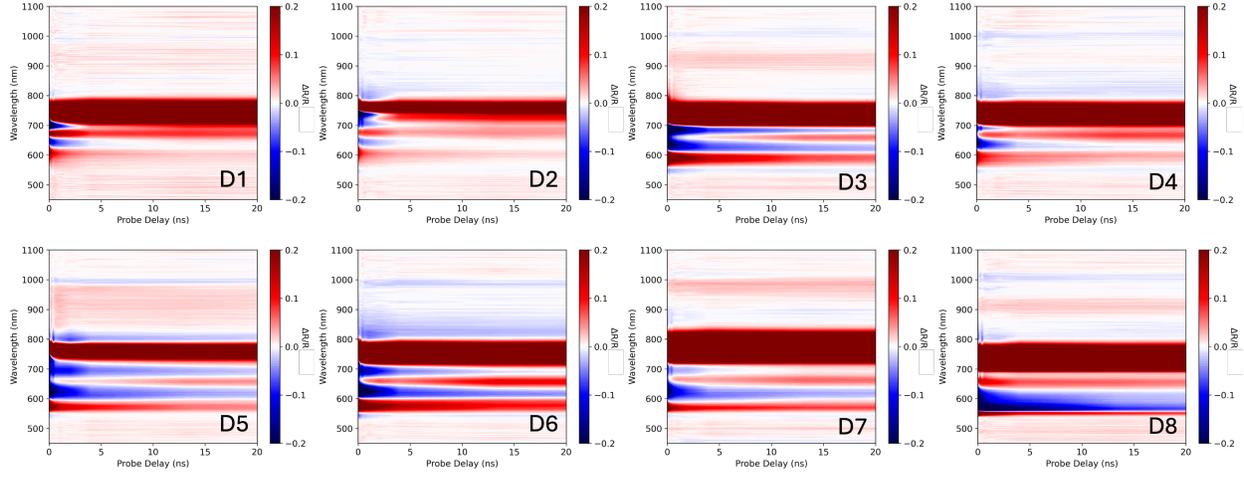

**Fig. S12**. **Nanosecond time scale dynamics of quantum battery devices.** Experimental pump probe differential reflectivity of each quantum battery device as a function of time up to 20 ns using the pump configuration specified in fig. S10.

### E. Theoretical model of ultrafast quantum battery charging

We developed a theoretical model to extract pertinent thermodynamic observables from the quantum battery ultrafast charging experiments such as, (i) the rise time $\tau$, (ii) the time-dependent energy density $E(t)$ and (iii) the time-dependent charging power $P(t)$ of each device. The coherent dynamics of the driven quantum battery is governed by the Dicke Hamiltonian

$$H(t) = (\Delta_c - \nu) a^\dagger a + \sum_{j=1}^{N} \left( (\Delta_1 - \nu) X_{11}^{(j)} + (\Delta_2 - \nu) X_{22}^{(j)} + \Delta_T X_{TT}^{(j)} \right)$$
$$+ g \sum_{j=1}^{N} \left( a^\dagger \left( X_{01}^{(j)} + X_{02}^{(j)} \right) + \left( X_{10}^{(j)} + X_{20}^{(j)} \right) a \right) + i\eta(t) \left( a^\dagger - a \right) \quad (4)$$

written in the rotating frame of the laser. Contrary to the coupled oscillator model, the Hilbert operators $X_{\alpha\beta}^{(j)} = |\alpha_j\rangle \langle \beta_j|$ now span the ground, excited Davydov states and the CuPc triplet state ($\Delta_T = 1.2$ eV) of each molecule (indexed by $j$) i.e. $\alpha_j, \beta_j \in \{0, 1, 2, T\}$. Since the light-matter interaction strength $g \propto \sqrt{\Delta_c/2\epsilon_0 V}$ depends on the volume $V$ occupied by the pumped CuPc molecules, we must account for a change in $g$ that results from changing the spot size from the steady state reflectance experiments to the much smaller spot size used in the ultrafast measurements. We find for our ultrafast experiments that $g \rightarrow g\sqrt{V_{SS}/V_{UF}} \approx 500$ neV where $V_{SS/UF}$ is the volume cross-section of the light beam in the steady-state/ultra-fast reflectance studies. For the pulse envelope $\eta(t)$, we assume a Gaussian lineshape of the form $\eta(t) = \sqrt{\frac{rN}{2\pi\sigma^2}} \exp\left(\frac{-(t-t_0)^2}{2\sigma}\right)$ where $r = \frac{1}{2}$ and $\sigma = \frac{35}{2\sqrt{2\log 2}}$ fs is obtained from the autocorrelation studies in figs. S6 and S7.

We account for the incoherent dynamics of the quantum battery coupled to a dissipative environment via a Lindblad master equation of the form

$$\dot{\rho} = -[H(t), \rho] + \kappa \mathcal{L}[a] + \sum_{j=1}^{N} \left( \gamma^- \left( \mathcal{L}[X_{01}^{(j)}] + \mathcal{L}[X_{02}^{(j)}] \right) + \gamma_T^- \mathcal{L}[X_{0T}^{(j)}] \right.$$
$$\left. + \gamma^z \left( \mathcal{L}[X_{11}^{(j)}] + \mathcal{L}[X_{22}^{(j)}] \right) + \gamma_{\text{ISC}} \left( \mathcal{L}[X_{T1}^{(j)}] + \mathcal{L}[X_{T2}^{(j)}] \right) \right) \quad (5)$$

where $\rho$ is the reduced density matrix of the cavity quantum battery and $\mathcal{L}[O] = O\rho O^\dagger - \frac{1}{2}\{O^\dagger O, \rho\}$ are Lindblad jump operators for an operator $O$. In Eq. (5) we account for cavity loss $\kappa$, non-radiative relaxation from the Davydov-split excited singlet state $\gamma^-$, non-radiative relaxation from the CuPc triplet state $\gamma_T^-$, dephasing of the molecular ensemble $\gamma^z$ and intersystem crossing from the excited singlet to triplet states $\gamma_{ISC}$.

Making use of $\frac{d\langle O \rangle}{dt} = \text{Tr}\{O\dot{\rho}\}$, we derive equations of motion for the photon, matter and mixed photon-matter operator expectation values for the quantum battery open quantum system[2]. The time evolution of each Hilbert operator represents an averaged time-evolution across the molecular ensemble i.e. $\langle X_{\alpha\beta} \rangle = \frac{1}{N} \sum_{j=1}^{N} \langle X_{\alpha\beta}^{(j)} \rangle$ and $\langle X_{\alpha\beta} X_{\sigma\tau} \rangle = \frac{1}{N(N-1)} \sum_{i \neq j}^{N} \langle X_{\alpha\beta}^{(i)} X_{\sigma\tau}^{(j)} \rangle$. Thus to ease notation, we omit molecular labels $j$ that index these operators in the following equations of motion. The equations of motion for the photon field confined to the cavity are

$$\frac{d\langle a \rangle}{dt} = -\left(\frac{\kappa}{2} + i(\Delta_c - \nu)\right)\langle a \rangle - iNg\langle X_{01} \rangle - iNg\langle X_{02} \rangle + \eta(t)$$

$$\frac{d\langle a^\dagger a \rangle}{dt} = -\kappa \langle a^\dagger a \rangle + 2\eta(t)\,\text{Re}\langle a^\dagger \rangle + 2gN\left(\text{Im}\langle a^\dagger X_{01}\rangle + \text{Im}\langle a^\dagger X_{02}\rangle\right) \quad (6)$$

$$\frac{d\langle aa \rangle}{dt} = -(\kappa + 2i(\Delta_c - \nu))\langle a \rangle - 2iNg\langle aX_{01}\rangle - iNg\langle aX_{02}\rangle + 2\eta(t)\langle a \rangle.$$

The time-evolution of the unimolecular expectation values is governed by

$$\frac{d\langle X_{\alpha\beta} \rangle}{dt} = -i(\Delta_1 - \nu)(\langle X_{\alpha 1}\rangle\delta_{1\beta} - \langle X_{1\beta}\rangle\delta_{\alpha 1}) - i(\Delta_2 - \nu)(\langle X_{\alpha 2}\rangle\delta_{2\beta} - \langle X_{2\beta}\rangle\delta_{\alpha 2})$$
$$- i\Delta_T(\langle X_{\alpha T}\rangle\delta_{T\beta} - \langle X_{T\beta}\rangle\delta_{\alpha T}) - ig\left(\langle a^\dagger X_{\alpha 1}\rangle\delta_{0\beta} - \langle a^\dagger X_{0\beta}\rangle\delta_{\alpha 1} + \langle aX_{\alpha 0}\rangle\delta_{1\beta} - \langle aX_{1\beta}\rangle\delta_{\alpha 0}\right.$$
$$+\langle a^\dagger X_{\alpha 2}\rangle\delta_{0\beta} - \langle a^\dagger X_{0\beta}\rangle\delta_{\alpha 2} + \langle aX_{\alpha 0}\rangle\delta_{2\beta} - \langle aX_{2\beta}\rangle\delta_{\alpha 0}\Big) + \gamma^-\left(\mathcal{R}_{\alpha\beta 01} + \mathcal{R}_{\alpha\beta 02}\right) + \gamma_T^-\mathcal{R}_{\alpha\beta 0T}$$
$$+ \gamma^z \left(\mathcal{R}_{\alpha\beta 11} + \mathcal{R}_{\alpha\beta 22}\right) + \gamma_{\text{ISC}}\left(\mathcal{R}_{\alpha\beta T1} + \mathcal{R}_{\alpha\beta T2}\right)$$

$$(7)$$

where $\mathcal{R}_{\alpha\beta\sigma\tau} = \langle X_{\tau\tau}\rangle\delta_{\alpha\sigma}\delta_{\beta\sigma} - \frac{1}{2}\langle X_{\tau\beta}\rangle\delta_{\alpha\tau} - \frac{1}{2}\langle X_{\alpha\tau}\rangle\delta_{\beta\tau}$ and $\delta_{\alpha\beta}$ is the Kronecker delta function. The equations of motion governing the time-evolution of the bimolecular expectation values

are

$$\frac{d\langle X_{\alpha\beta}X_{\sigma\tau}\rangle}{dt} = -i\left(\Delta_1 - \nu\right)\left(\langle X_{\alpha\beta}X_{\sigma 1}\rangle\delta_{\tau 1} - \langle X_{\alpha\beta}X_{1\tau}\rangle\delta_{\sigma 1} + \langle X_{\alpha 1}X_{\sigma\tau}\rangle\delta_{\beta 1} - \langle X_{1\beta}X_{\sigma\tau}\rangle\delta_{\alpha 1}\right)$$

$$- i\left(\Delta_2 - \nu\right)\left(\langle X_{\alpha\beta}X_{\sigma 2}\rangle\delta_{\tau 2} - \langle X_{\alpha\beta}X_{2\tau}\rangle\delta_{\sigma 2} + \langle X_{\alpha 2}X_{\sigma\tau}\rangle\delta_{\beta 2} - \langle X_{2\beta}X_{\sigma\tau}\rangle\delta_{\alpha 2}\right)$$

$$- i\Delta_T\left(\langle X_{\alpha\beta}X_{\sigma T}\rangle\delta_{\tau T} - \langle X_{\alpha\beta}X_{T\tau}\rangle\delta_{\sigma T} + \langle X_{\alpha T}X_{\sigma\tau}\rangle\delta_{\beta T} - \langle X_{T\beta}X_{\sigma\tau}\rangle\delta_{\alpha T}\right)$$

$$- ig\left(\left(\langle a^\dagger X_{\alpha\beta}X_{\sigma 1}\rangle + \langle a^\dagger X_{\alpha\beta}X_{\sigma 2}\rangle\right)\delta_{\tau 0} - \langle a^\dagger X_{\alpha\beta}X_{0\tau}\rangle\left(\delta_{\sigma 1} + \delta_{\sigma 2}\right) + \left(\langle a^\dagger X_{\alpha 1}X_{\sigma\tau}\rangle\right.\right.$$

$$\left.+\langle a^\dagger X_{\alpha 2}X_{\sigma\tau}\rangle\right)\delta_{\beta 0} - \langle a^\dagger X_{0\beta}X_{\sigma\tau}\rangle\left(\delta_{\alpha 1} + \delta_{\alpha 2}\right) + \langle aX_{\alpha\beta}X_{\sigma 0}\rangle\left(\delta_{\tau 1} + \delta_{\tau 2}\right)$$

$$- \left(\langle aX_{\alpha\beta}X_{1\tau}\rangle + \langle aX_{\alpha\beta}X_{2\tau}\rangle\right)\delta_{\sigma 0} + \langle aX_{\alpha 0}X_{\sigma\tau}\rangle\left(\delta_{\beta 1} + \delta_{\beta 2}\right) + \left(\langle aX_{\alpha 0}X_{\sigma\tau}\rangle\right.$$

$$\left.\langle aX_{1\beta}X_{\sigma\tau}\rangle\right)\delta_{\alpha 0} + \gamma_T^- \mathcal{S}_{\alpha\beta\sigma\tau 0T} + \gamma^-\left(\mathcal{S}_{\alpha\beta\sigma\tau 01} + \mathcal{S}_{\alpha\beta\sigma\tau 02}\right) + \gamma^z\left(\mathcal{S}_{\alpha\beta\sigma\tau 11} + \mathcal{S}_{\alpha\beta\sigma\tau 22}\right)$$

$$+ \gamma_{\text{ISC}}\left(\mathcal{S}_{\alpha\beta\sigma\tau T1} + \mathcal{S}_{\alpha\beta\sigma\tau T2}\right) \tag{8}$$

where $\mathcal{S}_{\alpha\beta\sigma\tau\xi\zeta} = \langle X_{\alpha\beta}X_{\zeta\zeta}\rangle\delta_{\sigma\xi}\delta_{\tau\xi} + \langle X_{\zeta\zeta}X_{\sigma\tau}\rangle\delta_{\alpha\xi}\delta_{\beta\xi} - \frac{1}{2}\langle X_{\alpha\beta}X_{\zeta\tau}\rangle\delta_{\sigma\zeta} - \frac{1}{2}\langle X_{\alpha\beta}X_{\sigma\zeta}\rangle\delta_{\tau\zeta} - \frac{1}{2}\langle X_{\zeta\beta}X_{\sigma\tau}\rangle\delta_{\alpha\zeta} - \frac{1}{2}\langle X_{\alpha\zeta}X_{\sigma\tau}\rangle\delta_{\beta\zeta}$ describes the dissipative dynamics of the two-body correlators and $\alpha,\beta,\sigma,\tau \in \{0,1,2,T\}$. The time-evolution of the mixed light-matter expectation values are obtained via the equations of motion

$$\frac{d\langle aX_{\alpha\beta}\rangle}{dt} = -\left(\frac{\kappa}{2} + i\left(\Delta_c - \nu\right)\right)\langle aX_{\alpha\beta}\rangle + \eta(t)\langle X_{\alpha\beta}\rangle - i\left(\Delta_1 - \nu\right)\left(\langle aX_{\alpha 1}\rangle\delta_{1\beta} - \langle aX_{1\beta}\rangle\delta_{\alpha 1}\right)$$

$$- i\left(\Delta_2 - \nu\right)\left(\langle aX_{\alpha 2}\rangle\delta_{2\beta} - \langle aX_{2\beta}\rangle\delta_{\alpha 2}\right) - i\Delta_T\left(\langle aX_{\alpha T}\rangle\delta_{T\beta} - \langle aX_{T\beta}\rangle\delta_{\alpha T}\right) - ig\left(\langle X_{\alpha 1}\rangle\delta_{\beta 0}\right.$$

$$\left.+\langle X_{\alpha 2}\rangle\delta_{\beta 0} + (N-1)\left(\langle X_{\alpha\beta}X_{01}\rangle + \langle X_{\alpha\beta}X_{02}\rangle\right) + \left(\langle a^\dagger aX_{\alpha 1}\rangle + \langle a^\dagger aX_{\alpha 2}\rangle\right)\delta_{\beta 0}\right.$$

$$\left.+ \left(\langle aaX_{1\beta}\rangle + \langle aaX_{2\beta}\rangle\right)\delta_{\alpha 0} + \langle a^\dagger aX_{0\beta}\rangle\left(\delta_{\alpha 1} + \delta_{\alpha 2}\right) + \langle aaX_{\alpha 0}\rangle\left(\delta_{1\beta} + \delta_{2\beta}\right)\right)$$

$$+ \gamma^-\left(\mathcal{P}_{\alpha\beta 01} + \mathcal{P}_{\alpha\beta 02}\right) + \gamma_T^- \mathcal{P}_{\alpha\beta 0T} + \gamma^z\left(\mathcal{P}_{\alpha\beta 11} + \mathcal{P}_{\alpha\beta 22}\right) + \gamma_{\text{ISC}}\left(\mathcal{P}_{\alpha\beta T1} + \mathcal{P}_{\alpha\beta T2}\right) \tag{9}$$

and

$$\begin{aligned}
\frac{d\langle a^\dagger X_{\alpha\beta}\rangle}{dt} = & -\left(\frac{\kappa}{2} - i(\Delta_c - \nu)\right)\langle a^\dagger X_{\alpha\beta}\rangle + \eta(t)\langle X_{\alpha\beta}\rangle - i(\Delta_1 - \nu)\left(\langle a^\dagger X_{\alpha 1}\rangle\delta_{1\beta} - \langle a^\dagger X_{1\beta}\rangle\delta_{\alpha 1}\right) \\
& - i(\Delta_2 - \nu)\left(\langle a^\dagger X_{\alpha 2}\rangle\delta_{2\beta} - \langle a^\dagger X_{2\beta}\rangle\delta_{\alpha 2}\right) - i\Delta_T\left(\langle a^\dagger X_{\alpha T}\rangle\delta_{T\beta} - \langle a^\dagger X_{T\beta}\rangle\delta_{\alpha T}\right) - ig\left(\langle X_{1\beta}\rangle\delta_{\alpha 0}\right. \\
& + \langle X_{2\beta}\rangle\delta_{\alpha 0} + (N-1)\left(\langle X_{\alpha\beta}X_{10}\rangle + \langle X_{\alpha\beta}X_{20}\rangle\right) + \left(\langle a^\dagger a X_{1\beta}\rangle + \langle a^\dagger a X_{2\beta}\rangle\right)\delta_{\alpha 0} \\
& + \left(\langle a^\dagger a^\dagger X_{\alpha 1}\rangle + \langle a^\dagger a^\dagger X_{\alpha 2}\rangle\right)\delta_{\beta 0} + \langle a^\dagger a X_{\alpha 0}\rangle\left(\delta_{\beta 1} + \delta_{\beta 2}\right) + \left.\langle a^\dagger a^\dagger X_{0\beta}\rangle\left(\delta_{\alpha 1} + \delta_{\alpha 2}\right)\right) \\
& + \gamma^-\left(\mathcal{Q}_{\alpha\beta 01} + \mathcal{Q}_{\alpha\beta 02}\right) + \gamma_T^-\mathcal{Q}_{\alpha\beta 0T} + \gamma^z\left(\mathcal{Q}_{\alpha\beta 11} + \mathcal{Q}_{\alpha\beta 22}\right) + \gamma_{\text{ISC}}\left(\mathcal{Q}_{\alpha\beta T1} + \mathcal{Q}_{\alpha\beta T2}\right)
\end{aligned} \tag{10}$$

where $\mathcal{P}_{\alpha\beta\sigma\tau} = \langle a X_{\tau\tau}\rangle\delta_{\alpha\sigma}\delta_{\beta\sigma} - \frac{1}{2}\langle a X_{\tau\beta}\rangle\delta_{\alpha\tau} - \frac{1}{2}\langle a X_{\alpha\tau}\rangle\delta_{\beta\tau}$ and $\mathcal{Q}_{\alpha\beta\sigma\tau} = \langle a^\dagger X_{\tau\tau}\rangle\delta_{\alpha\sigma}\delta_{\beta\sigma} - \frac{1}{2}\langle a^\dagger X_{\tau\beta}\rangle\delta_{\alpha\tau} - \frac{1}{2}\langle a^\dagger X_{\alpha\tau}\rangle\delta_{\beta\tau}$. Notably, the time evolution of the matter and mixed light-matter correlation functions depend upon three body correlation functions. To capture the relevant physical processes whilst also closing these equations of motion, we invoke the approximation

$$\langle\langle MNO\rangle\rangle = \langle MNO\rangle - \langle MN\rangle\langle O\rangle - \langle M\rangle\langle NO\rangle - \langle MO\rangle\langle N\rangle + 2\langle M\rangle\langle N\rangle\langle O\rangle \approx 0 \tag{11}$$

in which the third order cumulants vanish. At this order of approximation, errors scale as $O(1/N)$ and thus become insignificant for our simulations where $N \sim 10^{10}{}^2$.

After numerical integration of the above equations of motion, the experimental differential reflectance $\Delta R/R$ was simulated (fig. S13) by convolving the time-evolution of the sum of the excited CuPc absorber populations $\langle X_{11}\rangle + \langle X_{22}\rangle + \langle X_{TT}\rangle$ with an instrument response function, approximated as a Gaussian lineshape with full-width at half-maximum 50 fs. In addition, we calculated the time-evolution of the energy density $E(t) = \Delta_1\langle X_{11}\rangle + \Delta_2\langle X_{22}\rangle + \Delta_T\langle X_{TT}\rangle$ the charging time of each device $\tau$ (defined by $E(\tau) = E_{\max}/2$) for each device and the maximum charging power density $P_{\max} = \frac{E_{\max}}{\tau}$ (plotted in fig. 2 of the main text).

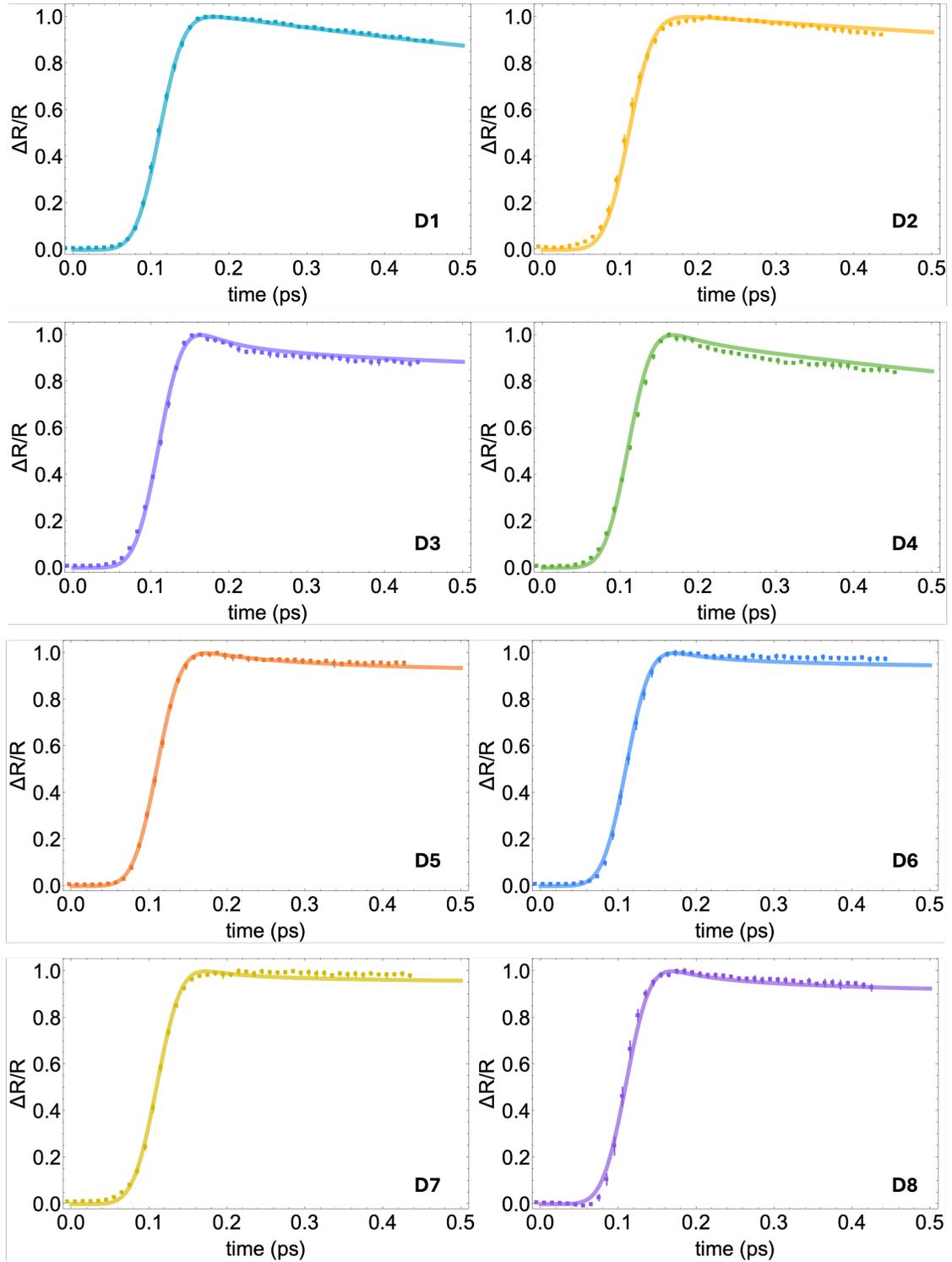

**Fig. S13**. **Simulated differential reflectance spectra of the quantum battery devices.** The simulation (using parameters from Table S3) are shown as a solid line superimposed with the experimental data (squares with error bars).

| Device | $N(\times 10^{10})$ | $\kappa$ (ps$^{-1}$) | $\gamma^-$ (ps$^{-1}$) | $\gamma_T^-$ (ps$^{-1}$) | $\gamma^z$ (ps$^{-1}$) | $\gamma_{\text{ISC}}$ (ps$^{-1}$) |
|---|---|---|---|---|---|---|
| D1 | 2.2 | 25 | 0.5 | 0.4 | 20 | 5 |
| D2 | 2.6 | 25 | 0.3 | 0.2 | 20 | 5 |
| D3 | 3.0 | 33 | 0.5 | 0.1 | 17 | 5 |
| D4 | 3.8 | 29 | 0.5 | 0.4 | 17 | 5 |
| D5 | 5.0 | 29 | 0.5 | 0.01 | 20 | 5 |
| D6 | 5.6 | 33 | 0.5 | 0.01 | 20 | 5 |
| D7 | 6.1 | 33 | 0.5 | 0.01 | 20 | 5 |
| D8 | 6.2 | 40 | 0.5 | 0.01 | 20 | 5 |

**Table. S3**. **Ultrafast quantum battery charging parameters.** For all simulations the $\Delta_1$, $\Delta_2$ and $\Delta_c$ were obtained from the coupled oscillator model (see Table S1) and $\Delta_T = 1.2$ eV. The bare light-matter coupling $g$ was fixed at 500 neV and the laser frequency was tuned to the lower polariton energy level of each device.

## IV. DEVICE VOLTAGE-CURRENT-POWER CHARACTERISTICS

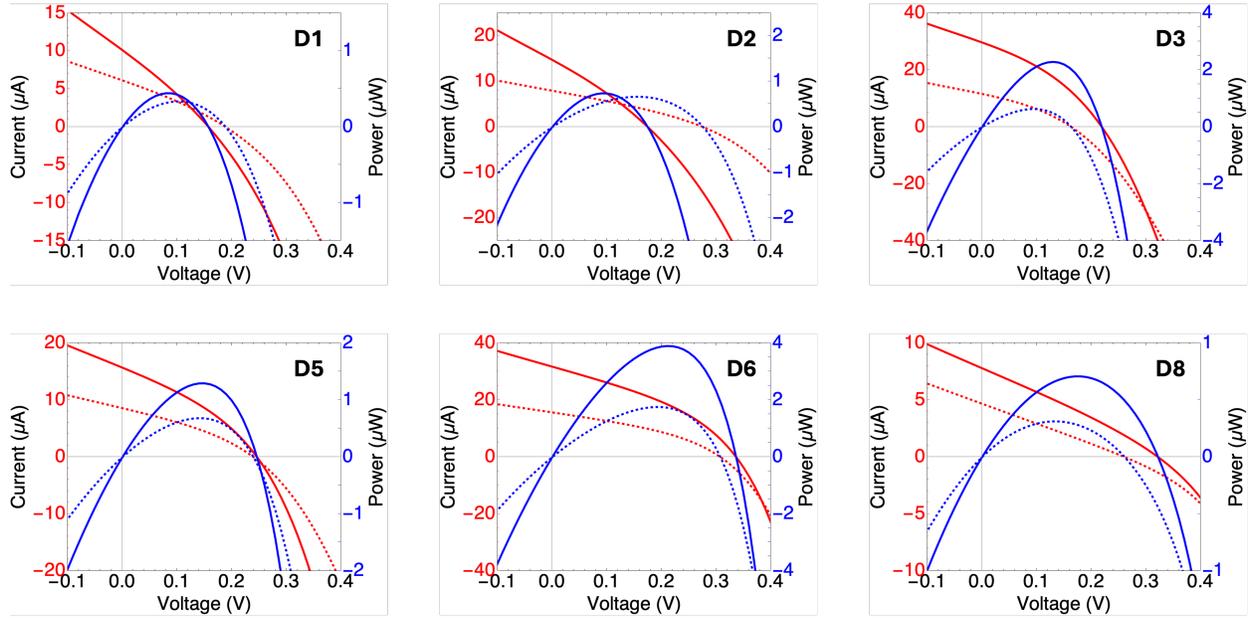

Fig. S14. **Voltage-current-power measurements for each device.** The steady-state current (red) and discharging power (blue) as a function of voltage for the cavity (solid lines) and no-cavity control (dashed line) are shown as function of voltage. Note that D4 and D7 are not reported here as these devices were fabricated without electrical controls and so it was impossible to measure their electrical response to compare with the cavity quantum battery.